\documentclass[10pt,aps,prx,twocolumn,notitlepage,showpacs,superscriptaddress]{revtex4-1}
\usepackage{amsthm,amsmath,amsfonts,amssymb,verbatim, color}
\usepackage{graphicx}
\usepackage{bm}
\usepackage{epsfig,slashed}
\usepackage[T1]{fontenc}
\usepackage[colorlinks=true,citecolor=blue,linkcolor=blue,urlcolor=blue]{hyperref}
\usepackage{amsmath}
\usepackage{enumitem}

\newcommand{\im}{\mathop{\mathrm{Im}}}

\renewcommand{\b}[1]{{\boldsymbol{#1}}}
\renewcommand{\mathbf}[1]{{\boldsymbol{#1}}}

\renewcommand{\c}[1]{\mathcal{#1}}

\newcommand{\upa}{\uparrow}
\newcommand{\dwa}{\downarrow}

\newcommand{\nn}{\nonumber}

\newcommand{\Fig}[1]{Fig.~\ref{#1}} 
\newcommand{\eq}[1]{Eq.~(\ref{#1})} 
\newcommand{\mean}[1]{\langle#1\rangle}

\newcommand{\da}[1]{d^{}_{#1}}
\newcommand{\dc}[1]{d^{\dagger}_{#1}}

\begin{document}

\title{Gapless helical superconductivity on the surface of a three-dimensional topological insulator}

\author{Isil Ozfidan}
\affiliation{Department of Physics, University of Alberta, Edmonton, Alberta T6G 2E1, Canada}

\author{Jinsen Han}
\affiliation{Department of Physics, University of Alberta, Edmonton, Alberta T6G 2E1, Canada}
\affiliation{Department of Physics, College of Science, National University of Defense Technology, Changsha 410073, P. R. China}

\author{Joseph Maciejko}
\affiliation{Department of Physics, University of Alberta, Edmonton, Alberta T6G 2E1, Canada}
\affiliation{Theoretical Physics Institute, University of Alberta, Edmonton, Alberta T6G 2E1, Canada}
\affiliation{Canadian Institute for Advanced Research, Toronto, Ontario M5G 1Z8, Canada}

\date\today

\begin{abstract}
Recent angle-resolved photoemission experiments have observed a proximity-induced superconducting gap in the helical surface states of a thin film of the 3D topological insulator Bi$_2$Se$_3$ grown on a superconducting NbSe$_2$ substrate. The superconducting coherence peaks in the electronic density of states are strongly suppressed when the topological insulator is doped with magnetic Mn impurities, which was interpreted as the complete destruction of helical superconductivity in the topological surface states. Motivated by these experiments, we explore a different possibility: gapless helical superconductivity, where a gapless electronic density of states coexists with a nonzero helical superconducting order parameter. We study a model of superconducting Dirac fermions coupled to random magnetic impurities within the Abrikosov-Gor'kov framework, and find finite regions of gapless helical superconductivity in the phase diagram of the system for both proximity-induced and intrinsic superconductivity. For the latter, we derive universal rates of supression of the superconducting transition temperature due to magnetic scattering and, for a Fermi level at the Dirac point, a universal rate of increase of the quantum critical attraction strength.
\end{abstract}


\maketitle

\section{Introduction}

The interplay between topological and conventional forms of order is a central theme of current research in condensed matter physics. In particular, superconductivity on the 2D boundary of a 3D topological insulator, sometimes referred to as helical superconductivity, is qualitatively distinct from conventional 2D $s$-wave superconductivity. By contrast with the latter, vortices in a helical superconductor are predicted to support Majorana fermions~\cite{fu2008}, and the quantum phase transition between a semimetal and a helical superconductor is conjectured to exhibit an emergent supersymmetry~\cite{Grover14,Lee14,grover2012}. On the experimental side, tremendous effort has been expended in the past few years to induce helical superconductivity in the topological surface states. The observation of a proximity-induced superconducting gap has been reported in Bi$_2$Se$_3$ topological insulator films grown on conventional $s$-wave superconducting substrates such as W~\cite{zhang2011}, NbSe$_2$~\cite{wang2012,xu2014}, and Sn~\cite{yang2012}, as well as on a high-temperature $d$-wave superconducting Bi$_2$Sr$_2$CaCu$_2$O$_{8+\delta}$ substrate~\cite{wang2013}. Observations of a Josephson supercurrent flowing through the topological surface state have been reported in a variety of topological insulator-superconductor junctions involving the topological insulator materials Bi$_2$Se$_3$~\cite{sacepe2011,williams2012,cho2013,kurter2014,finck2014,stehno2016,finck2016}, Bi$_2$Te$_3$~\cite{veldhorst2012,qu2012,xu2014b}, HgTe~\cite{maier2012,oostinga2013,sochnikov2015,wiedenmann2016}, and Bi$_{1.5}$Sb$_{0.5}$Te$_{1.7}$Se$_{1.3}$~\cite{snelder2015}. There is also evidence that intrinsic, as opposed to proximity-induced, superconductivity may coexist with topological surface states in the topological insulator SbTe$_3$~\cite{zhu2013,zhao2015} and the layered superconductor $\beta$-PdBi$_2$~\cite{sakano2015}. In a recent experiment~\cite{xu2014}, spin- and angle-resolved photoemission spectroscopy (ARPES) measurements provided essentially direct experimental evidence for proximity-induced helical superconductivity in Bi$_2$Se$_3$ thin films on a superconducting NbSe$_2$ substrate. A unique, spin-momentum-locked Fermi surface was observed above the superconducting transition temperature of NbSe$_2$ ($T_c=7.2$~K), and below $T_c$ a relatively isotropic spectral gap (leading-edge shifts) of $\sim$0.7~meV and coherence peaks were seen to form. 

An important question in this context is the role of time-reversal-symmetry-breaking perturbations on helical superconductivity. One expects such perturbations to be particularly detrimental to this type of superconductivity for two reasons. First, the topological surface states owe their very existence to time-reversal symmetry, which protects the bulk topology of the normal state. Time-reversal-symmetry-breaking perturbations can in principle gap out the surface states and destroy the parent metallic state necessary for helical superconductivity. Second, superconductivity itself is triggered by the formation of Cooper pairs, in which electrons occupy two time-reversed single-particle eigenstates that are degenerate by Kramers' theorem. While the latter holds in the presence of time-reversal invariant forms of disorder, such as nonmagnetic impurities, time-reversal-symmetry-breaking forms of disorder such as magnetic impurities destroy Kramers' degeneracy and suppress superconductivity~\cite{anderson1959}. In the ARPES experiment mentioned earlier~\cite{xu2014}, the superconducting coherence peaks were found to be strongly suppressed upon doping the topological Bi$_2$Se$_3$ layer with magnetic Mn impurities (4\% and 10\% concentration). This was interpreted as the complete suppression of superconductivity in the topological surface states, with the magnetic doping driving a phase transition from a helical superconductor to a normal Dirac metallic state.

In this paper we suggest another possible scenario: that in certain parameter regimes, magnetic impurities might induce a gapless superconductor in the topological surface states, rather than destroy superconductivity altogether. In this gapless helical superconductor, coherence peaks in the density of states are suppressed by disorder, but the helical superconducting order parameter itself (i.e., the presence of Cooper pairs) is nonzero. To demonstrate these ideas, we organize the rest of the paper as follows. In Sec.~\ref{sec:ReviewSCDirac}, we review the basic principles of superconductivity in the topological surface states in the clean limit, and introduce a theoretical framework to study the effect of magnetic impurities in this system. In Sec.~\ref{sec:prox} and \ref{sec:intrinsic}, respectively, we apply these ideas to separately investigate the effect of magnetic impurities on proximity-induced and intrinsic superconductivity. We briefly conclude in Sec.~\ref{sec:conclusion}.

\section{Helical superconductivity and magnetic impurities}
\label{sec:ReviewSCDirac}

\subsection{Clean limit}

In order to study the effects of magnetic impurities on superconducting surfaces of 3D topological insulators, we begin with a brief review of the Bardeen-Cooper-Schrieffer (BCS) theory of superconductivity applied to the Dirac surface states without any impurities. The effective Hamiltonian for noninteracting Dirac electrons on the surface is given by
\begin{equation}\label{H0}
H_0=\sum_{\mathbf{k}\sigma\sigma'} c^{\dagger}_{\b{k}\sigma}\left[v_F\hat{\b{z}}\cdot(\b{\sigma}\times\b{k})_{\sigma\sigma'} -\mu\delta_{\sigma\sigma'}\right] c_{\b{k}\sigma'}, 
\end{equation}
where $c^{\dagger}_{\b{k}\sigma}$ ($c_{\b{k}\sigma}$) creates (annihilates) an electron with spin $\sigma=\uparrow,\downarrow$ and momentum $\b{k}$, $\mu$ is the chemical potential, $v_F$ is the Dirac velocity, $\b{\sigma}$ is a vector of Pauli matrices representing spin, and $\hat{\b{z}}$ is a vector normal to the surface. The Hamiltonian (\ref{H0}) is diagonal in the helicity basis described by the operators $d_{\b{k}\pm}^\dagger$ that create electrons on the upper ($+$)/lower ($-$) branch of the Dirac spectrum,
\begin{equation}\label{helicity}
d_{\b{k}\eta}^{\dagger}=\frac{1}{\sqrt{2}}(c_{\b{k}\uparrow}^\dag +\eta e^{i\theta_\b{k}}c_{\b{k}\downarrow}^\dag),\hspace{5mm}\eta=\pm 1,
\end{equation}
where $e^{i\theta_\b{k}}=(k_y-ik_x)/|\b{k}|$. These operators create eigenstates of the noninteracting Hamiltonian (\ref{H0}) with energy eigenvalues $\pm\epsilon_\b{k}-\mu$ as measured from the chemical potential, where $\epsilon_\b{k}=v_F|\b{k}|$.

An attractive interaction leading to intrinsic superconductivity can be introduced via the reduced BCS interaction Hamiltonian,
\begin{align}\label{Hint}
H_\text{int}&=-g\sum_{\mathbf{k}\mathbf{k}'\sigma\sigma'}'
c^{\dagger}_{\b{k}\sigma}c_{\b{k}'\sigma}^{}c^{\dagger}_{-\b{k}\sigma'}c_{-\b{k}'\sigma'}^{}\nn\\
&=-2g\sum_{\mathbf{k}\mathbf{k}'}' c_{\b{k}\upa}^\dag c_{-\b{k}\dwa}^\dag c_{-\b{k}'\dwa}^{}c_{\b{k}'\upa}^{},
\end{align}
where $g$ is positive and only $s$-wave interactions in the Cooper channel are kept. In going from the first line to the second we have used Fermi statistics, which implies that fermions of the same spin cannot interact via a contact interaction. The prime on the sum signifies that the latter is restricted to momenta within a shell of width $\Lambda/v_F$ around the Fermi surface, where $\Lambda$ is a high-energy cutoff. Assuming that  attractive interactions are mediated by electron-phonon coupling, the cutoff $\Lambda$ would be given roughly by the Debye frequency $\omega_D$ of the phonons.

In this paper we will be interested in two separate limits: the limit of large chemical potential $\mu\gg\Lambda$, and the limit of zero chemical potential $\mu=0$. In the former only electrons on the upper branch of the Dirac spectrum pair, assuming $\mu>0$ without loss of generality, while in the latter there is pairing of electrons on both branches. To take both cases into account in a unified framework it turns out to be convenient to work in the basis of helicity eigenstates (\ref{helicity}). In the $\mu\gg\Lambda$ limit, we can discard the negative-helicity operators $d_{\b{k}-},d_{\b{k}-}^\dag$ entirely, and the interaction Hamiltonian (\ref{Hint}) becomes~\cite{ito2011,ito2012,nandkishore2013}
\begin{align}
H_\text{int}=-\frac{g}{2}\sum_{\b{k}\b{k}'}'\Bigl(e^{-i\theta_\b{k}}d_{\b{k}+}^\dag d_{-\b{k}+}^\dag\Bigr)
\Bigl(e^{i\theta_{\b{k}'}}d_{-\b{k}'+}^{}d_{\b{k}'+}^{}\Bigr),\nn\\
\mu\gg\Lambda.
\end{align}
Decoupling this interaction in the particle-particle channel, we obtain the mean-field interaction Hamiltonian
\begin{align}\label{HMFlargemu}
H_\text{int}^\text{MF}=-\frac{1}{2}\sum_\b{k}'\left(\Delta e^{-i\theta_\b{k}} d_{\b{k}+}^\dag d_{-\b{k}+}^\dag+\text{h.c.}\right),\hspace{5mm}\mu\gg\Lambda,
\end{align}
where the helical order parameter $\Delta$~\cite{fu2008} obeys the self-consistency condition
\begin{align}\label{GapEqlargemu}
\Delta=g\sum_{\b{k}}'e^{i\theta_{\b{k}}}\langle d_{-\b{k}+}d_{\b{k}+}\rangle_\text{MF},\hspace{5mm}\mu\gg\Lambda.
\end{align}
For zero chemical potential, only spin-singlet pairing is allowed~\cite{nandkishore2013}, and the interaction term (\ref{Hint}) is decoupled using the ansatz
\begin{eqnarray}
c^{\dagger}_{\b{k}\uparrow}c^{\dagger}_{-\b{k}\downarrow}c^{}_{-\b{k}'\downarrow}c^{}_{\b{k}'\uparrow}&\approx&
\mean{c^{\dagger}_{\b{k}\uparrow}c^{\dagger}_{-\b{k}\downarrow}}c^{}_{-\b{k}'\downarrow}c^{}_{\b{k}'\uparrow} \nonumber \\
&&+\mean{c^{}_{-\b{k}'\downarrow}c^{}_{\b{k}'\uparrow}}c^{\dagger}_{\b{k}\uparrow}c^{\dagger}_{-\b{k}\downarrow}.
\end{eqnarray}
However, fermions of both helicities must be kept. In the helicity basis, the mean-field interaction Hamiltonian becomes
\begin{equation}\label{HMFmu0}
H_\text{int}^\text{MF}=-\frac{1}{2}\sum_{\b{k}\eta}'\left(\Delta e^{-i\theta_\b{k}}\eta\dc{\b{k}\eta}\dc{-\b{k}\eta}+\text{h.c.}\right),\hspace{5mm}\mu=0,
\end{equation}
where the self-consistency condition obeyed by the order parameter $\Delta$ is
\begin{align}\label{gapeq}
\Delta=g\sum_{\b{k}\eta}'e^{i\theta_{\b{k}}}\eta\langle d_{-\b{k}\eta}d_{\b{k}\eta}\rangle_\text{MF},\hspace{5mm}\mu=0.
\end{align}
Using Eq.~(\ref{helicity}), we can show that Eq.~(\ref{gapeq}) is equivalent to the standard self-consistency condition for spin-singlet pairing, $\Delta=2g\sum_\b{k}\langle c_{\b{k}\upa}c_{-\b{k}\dwa}\rangle_\text{MF}$. This shows that pure spin-singlet pairing, relevant for $\mu=0$, is equivalent to a pairing amplitude that is equal in magnitude and opposite in sign for fermions of opposite helicity. Working in the helicity basis allows us to treat the $\mu\gg\Lambda$ and $\mu=0$ limits in a unified manner: the $\mu\gg\Lambda$ Eqs.~(\ref{HMFlargemu})-(\ref{GapEqlargemu}) correspond simply to dropping the negative-helicity parts of the $\mu=0$ Eqs.~(\ref{HMFmu0})-(\ref{gapeq}). Of course, the value of $\mu$ also enters the normal-state Hamiltonian (\ref{H0}). In both Eqs.~(\ref{GapEqlargemu}) and (\ref{gapeq}), the average $\langle\cdots\rangle_\text{MF}$ is taken in the grand canonical ensemble governed by the mean-field Hamiltonian
\begin{align}\label{HMF}
H_\text{MF}=H_0+H_\text{int}^\text{MF},
\end{align}
at temperature $T$ and chemical potential $\mu$. 

In the case of proximity-induced superconductivity, the parameter $\Delta$ is imposed by the bulk superconductor. Strictly speaking, the proximity effect causes $\Delta$ to be frequency-dependent, but this frequency dependence can be neglected if we are only interested in the behavior of the system for frequencies $\omega$ much less than the gap $\Delta_0$ of the bulk superconductor~\cite{sau2010,potter2011}. This is a self-consistent procedure if the proximity effect is weak $\Delta\ll\Delta_0$, and we are interested (as will be the case here) in frequencies that are on the order of $\Delta$ at the most.

\begin{figure}[t]
\begin{center}
\includegraphics[width=0.75\linewidth]{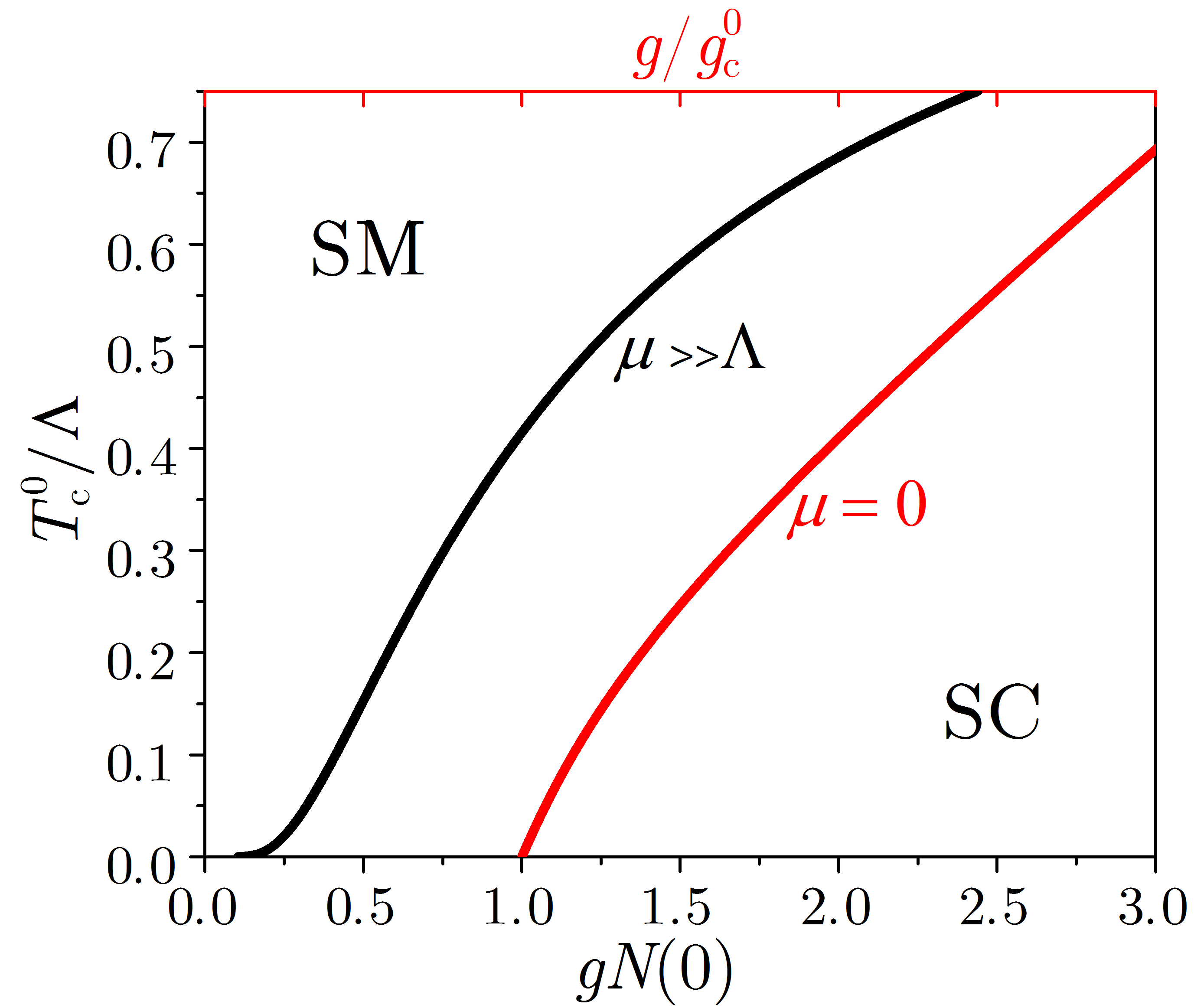}
\end{center}
\caption{Superconducting transition temperature $T_c^0$ of the clean system with chemical potential $\mu$, as a function of $gN(0)$ for $\mu\gg\Lambda$ (black) and $g/g_c^0$ for $\mu=0$ (red). SM: semimetal, SC: helical superconductor. Here $g$ is the BCS coupling, $\Lambda$ is the pairing energy scale (high-energy cutoff), $N(0)$ is the density of states at the Fermi level for nonzero $\mu$, and $g_c^0$ is the critical BCS coupling for $\mu=0$.} 
\label{gTclean}
\end{figure}

We now introduce Nambu spinors $\Psi_\b{k}$, in terms of which the mean-field Hamiltonian (\ref{HMF}) can be written as
\begin{align}
H_\text{MF}=\frac{1}{2}\sum_\b{k}\Psi^\dag_\b{k}H_\text{MF}(\b{k})\Psi_\b{k}^{},
\end{align}
where $H_\text{MF}(\b{k})$ is the mean-field Hamiltonian matrix. For $\mu\gg\Lambda$, a two-component Nambu spinor is sufficient,
\begin{equation}
\Psi_\b{k}=\left(d_{\b{k}+}^{},d_{-\b{k}+}^{\dagger}\right)^T,\hspace{5mm}\mu\gg\Lambda,
\end{equation}
and the Hamiltonian matrix is
\begin{align}\label{HBdGmuBig}
H_\text{MF}(\b{k})=(\epsilon_\b{k}-\mu)\tau_3-\Delta\b{\tau}\cdot\hat{\b{k}},\hspace{5mm}\mu\gg\Lambda,
\end{align}
where the Pauli matrices $\tau_{1,2,3}$ act on particle-hole indices and $\hat{\b{k}}=(\cos\theta_\b{k},\sin\theta_\b{k})=(k_y,-k_x)/|\b{k}|$ is a unit vector perpendicular to $\b{k}$. For $\mu=0$, we must keep fermions of both helicities and a four-component Nambu spinor is necessary,
\begin{equation}
\Psi_\b{k}=\left(d_{\b{k}+}^{},d_{-\b{k}+}^{\dagger},d_{\b{k}-}^{},d_{-\b{k}-}^{\dagger}\right)^T,\hspace{5mm}\mu=0.
\end{equation}
In this case, the mean-field Hamiltonian matrix is given by
\begin{align}\label{HBdGmu0}
H_\text{MF}(\b{k})=\epsilon_\b{k}\sigma_3\otimes\tau_3
-\Delta\sigma_3\otimes\b{\tau}\cdot\hat{\b{k}},\hspace{5mm}\mu=0,
\end{align}
where here the Pauli matrices $\sigma_{1,2,3}$ act on helicity indices. Finally, we introduce an imaginary-time Nambu Green's function
\begin{align}\label{G0}
\c{G}^{(0)}(\b{k},\tau)=-\mean{\mathcal{T}\Psi_\b{k}^{}(\tau)\Psi_\b{k}^\dag(0)},
\end{align}
where $\c{T}$ is the time-ordering operator in imaginary time and the superscript (0) signifies that the Green's function of the clean system (\ref{G0}) will act as the unperturbed Green's function for the system with disorder to be considered in the following sections. Fourier transforming to the fermionic Matsubara frequencies $\omega_n=(2n+1)\pi T$ where $T$ is the temperature and $n\in\mathbb{Z}$, the Green's function $\c{G}^{(0)}(\b{k},i\omega_n)$ is given by the solution of the Dyson equation,
\begin{align}\label{Ginvclean}
\c{G}^{(0)}(\b{k},i\omega_n)=(i\omega_n-H_\text{MF}(\b{k}))^{-1}.
\end{align}
The anomalous propagator or Gor'kov function $\mathcal{F}_{\eta}(\b{k},\tau)=-\mean{\mathcal{T}\da{\b{k}\eta}(\tau)\da{-\b{k}'\eta}(0)}$ for electrons with helicity $\eta$ can be extracted from the Nambu Green's function (\ref{G0}) as $\mathcal{F}_{+}(\b{k},\tau)=\c{G}^{(0)}_{12}(\b{k},\tau)$ and, for $\mu=0$, also as $\mathcal{F}_{-}(\b{k},\tau)=\c{G}^{(0)}_{34}(\b{k},\tau)$. These are then used in the self-consistency conditions (\ref{GapEqlargemu}) and (\ref{gapeq}). Calculating \eq{Ginvclean} explicitly, the self-consistency conditions (\ref{GapEqlargemu}) and (\ref{gapeq}) can be expressed as
\begin{equation}
1=gT\sum_\b{k}\sum_{i\omega_n}\frac{D(\mu)}{\omega_n^2+\xi_\b{k}^2+\Delta^2},
\label{gapeq2}
\end{equation} 
where $\xi_\b{k}=\epsilon_\b{k}-\mu$ and $D(\mu)$ is a multiplicity factor that takes on the values $D(\mu)=1$ for $\mu\gg\Lambda$ and $D(\mu)=2$ for $\mu=0$. This multiplicity factor appears due to the fact that fermions of only one helicity pair in the $\mu\gg\Lambda$ limit, while both pair with equal amplitude for $\mu=0$. The superconducting transition temperature $T_c^0$ for the clean system is obtained by setting the order parameter $\Delta$ to zero in Eq.~(\ref{gapeq2}) and performing the sum over Matsubara frequencies, which gives
\begin{align}\label{GapEquationE}
1=gD(\mu)\int_{-\Lambda}^\Lambda d\xi\,N(\xi)\frac{\tanh(|\xi|/2T_c^0)}{2|\xi|}.
\end{align}
For $\mu\gg\Lambda$, this gives the BCS-like expression
\begin{align}\label{mg0cleangt}
\frac{T_c^0}{\Lambda}\approx\frac{2e^\gamma}{\pi}e^{-1/gN(0)},\hspace{5mm}\mu\gg\Lambda,
\end{align}
where $N(0)$ is the density of states at the Fermi level and $\gamma\approx 0.577$ is Euler's constant. The black line in \Fig{gTclean} depicts $T_c$ as a function of $g$: there is a nonzero critical temperature for an arbitrarily weak attraction strength. For $\mu=0$ the density of states $N(\epsilon)=|\epsilon|/2\pi v_F^2$ vanishes at the Fermi level, and its energy dependence must be kept in Eq.~(\ref{GapEquationE}). We obtain the relation
\begin{equation}
    \frac{\pi v_F^2}{g}=2T_c^0\ln\cosh\left(\frac{\Lambda}{2T_c^0}\right),\hspace{5mm}\mu=0.
\end{equation}
The curve of $T_c^0$ versus $g$ deriving from this relation is plotted as a red line in \Fig{gTclean}. As is well-known for Dirac fermions, the critical temperature vanishes for a critical value $g_c^0$ of the BCS coupling~\cite{wilson1973,gross1974,zhao2006,kopnin2008,uchoa2009,roy2013,nandkishore2013},
\begin{align}\label{gc0}
g_c^0=\frac{\pi v_F^2}{\Lambda},
\end{align}
which corresponds to a quantum critical point between the Dirac semimetal for $g<g_c^0$ and the helical superconductor for $g>g_c^0$. There has been much interest in this quantum critical point recently, owing to the fact that it is predicted to display an emergent supersymmetry~\cite{grover2012,Grover14,Lee14,witczak-krempa2015}. In the present mean-field theory, the critical temperature vanishes linearly near the critical point,
\begin{align}\label{Tc0DiracPt}
\frac{T_c^0}{\Lambda}\approx\frac{1}{2\ln 2}\frac{g-g_c^0}{g_c^0},\hspace{5mm}\mu=0.
\end{align}
In general, near a quantum critical point with dynamic critical exponent $z=1$, one has $T_c^0\sim(g-g_c^0)^\nu$ where $\nu$ is the correlation length exponent~\cite{sachdev1997}; here mean-field theory predicts $\nu=1$, but fluctuations are known to reduce this value below one somewhat~\cite{Thomas05,Lee07,Bobev,zerf2016}.

\subsection{Magnetic impurities: Abrikosov-Gor'kov formalism}

Having established the properties of the clean system, we now introduce impurities. The scattering of electrons off a collection of $N_\text{imp}$ magnetic impurities at random positions $\{\b{R}_i\}$, $i=1,\ldots,N_\text{imp}$, can be described by the following exchange Hamiltonian
\begin{equation}\label{Himp}
H_\text{imp}=\frac{J}{V}\sum_{i=1}^{N_\text{imp}}\sum_{\b{k}\b{k}'\sigma\sigma'}e^{-i(\b{k}-\b{k}')\cdot \b{R}_i}\mathbf{S}_i\cdot c_{\b{k}\sigma}^{\dagger}\b{\sigma}_{\sigma\sigma'}c^{}_{\b{k}'\sigma'},
\end{equation}
where $V$ is the area of the topological insulator surface, $J$ is the exchange coupling between the magnetic impurities and the electron spin, and $\b{S}_i$ is the impurity spin. In the helicity basis, the Hamiltonian reads 
\begin{align}
H_\text{imp}=\frac{J}{2V}\sum_{i=1}^{N_\text{imp}}\sum_{\b{kk}'\eta\eta'}e^{-i(\b{k}-\b{k}')\cdot \b{R}_i}\mathbf{S}_i\cdot\dc{\b{k}\eta}\b{m}^{\b{k}\b{k}'}_{\eta\eta'}\da{\b{k}'\eta'},
\end{align}
where we define the vector $\b{m}^{\b{k}\b{k}'}=(m^{\b{k}\b{k}'}_x,m^{\b{k}\b{k}'}_y,m^{\b{k}\b{k}'}_z)$ of $2\times 2$ matrices with matrix elements
\begin{eqnarray}
(m^{\b{k}\b{k}'}_{\eta\eta'})_x&=&\eta e^{-i\theta_\b{k}} +\eta' e^{i\theta_{\b{k}'}}, \label{mx}\\
(m^{\b{k}\b{k}'}_{\eta\eta'})_y&=&i\eta e^{-i\theta_\b{k}} - i \eta' e^{i\theta_{\b{k}'}}, \\
(m^{\b{k}\b{k}'}_{\eta\eta'})_z&=&1-\eta\eta' e^{-i(\theta_\b{k}-\theta_{\b{k}'})}.\label{mz}
\end{eqnarray}
In the Nambu basis, we have 
\begin{align}\label{HimpNambu}
H_\text{imp}=\frac{1}{2}\sum_{\b{kk}'}\Psi_\b{k}^\dag\c{H}_\textrm{imp}(\b{k},\b{k}')\Psi_{\b{k}'},
\end{align}
where the Hamiltonian matrix $\c{H}_\textrm{imp}(\b{k},\b{k}')$ is
\begin{align}
\c{H}_\textrm{imp}(\b{k},\b{k}')=\frac{J}{2V}\sum_{i=1}^{N_\text{imp}}e^{-i(\b{k}-\b{k}')\cdot \b{R}_i}\mathbf{S}_i\cdot\b{\c{M}}^{\b{k}\b{k}'},
\end{align}
and we define the vector $\b{\c{M}}^{\b{k}\b{k}'}=(\c{M}^{\b{k}\b{k}'}_x,\c{M}^{\b{k}\b{k}'}_y,\c{M}^{\b{k}\b{k}'}_z)$ of Nambu matrices, 
\begin{align}\label{MNambu}
\b{\c{M}}^{\b{k}\b{k}'}=\left(\begin{array}{cccc}
\b{m}^{\b{kk}'}_{++} & 0 & \b{m}^{\b{kk}'}_{+-} & 0 \\
0 & -\b{m}^{-\b{k}',-\b{k}}_{++} & 0 & -\b{m}^{-\b{k}',-\b{k}}_{-+} \\
\b{m}^{\b{kk}'}_{-+} & 0 & \b{m}^{\b{kk}'}_{--} & 0 \\
0 & -\b{m}^{-\b{k}',-\b{k}}_{+-} & 0 & -\b{m}^{-\b{k}',-\b{k}}_{--}
\end{array}\right),
\end{align}
for $\mu=0$. For $\mu\gg\Lambda$, one only keeps the top left $2\times 2$ block of Eq.~(\ref{MNambu}). Hermiticity of the impurity Hamiltonian (\ref{HimpNambu}) requires $\b{\c{M}}^{\b{k}\b{k}'}=(\b{\c{M}}^{\b{k}'\b{k}})^\dag$, which is satisfied since $m_\alpha^{\b{kk}'}=(m_\alpha^{\b{k}'\b{k}})^\dag$, $\alpha=x,y,z$, as can be checked explicitly from Eq.~(\ref{mx})-(\ref{mz}).

To study the effect of magnetic impurity scattering on the superconducting properties of the system, we adapt the Abrikosov-Gor'kov formalism~\cite{abrikosov1961} to helical superconductivity. This was done previously in the $\mu\gg\Lambda$ limit only, for potential scattering~\cite{ito2011,ito2012,tkachov2013} and for scattering on polarized magnetic impurities~\cite{ito2012}, i.e., magnetic impurities whose positions are random but whose spin orientation is fixed. Here we will consider both the $\mu\gg\Lambda$ and $\mu=0$ limits, and consider magnetic impurities whose positions and spin orientations are both random. To study the $\mu\gg\Lambda$ limit, we simply omit all terms involving a negative helicity.

Treating impurity scattering perturbatively, the impurity-averaged Green's function $\c{G}(\b{k},i\omega_n)$ obeys a translationally invariant Dyson equation
\begin{equation}
\c{G}(\b{k},i\omega_n)^{-1}=\c{G}^{(0)}(\b{k},i\omega_n)^{-1}-\Sigma(\b{k},i\omega_n),
\label{dyson}
\end{equation}
where the self-energy $\Sigma(\b{k},i\omega_n)$ can be computed order by order in the exchange coupling $J$ and the impurity concentration $n_\text{imp}=N_\text{imp}/V$. To leading order in $n_\text{imp}$, we have
\begin{align}
\frac{1}{V}\left\langle e^{-i\b{q}\cdot\b{R}_i} e^{-i\b{q}'\cdot\b{R}_j}\right\rangle_\textrm{imp}=n_\textrm{imp}\delta_{ij}\delta_{\b{q}+\b{q}',0}.
\end{align}
Because the spin orientation of the impurities is random, we also have
\begin{align}
\left\langle S_i^\alpha S_j^\beta \right\rangle_\text{imp}=\frac{1}{3}S(S+1)\delta_{ij}\delta_{\alpha\beta},
\end{align}
where $S$ is the spin of the impurity. To first order in $J$, the self-energy is proportional to the impurity average of a single spin operator, which vanishes. To second order in $J$, the self-energy is given by
\begin{align}\label{Sigma2}
&\Sigma^{(2)}(\b{k},i\omega_n)=\sum_\b{p}\left\langle \c{H}_\text{imp}(\b{k},\b{p})\c{G}^{(0)}(\b{p},i\omega_n)\c{H}_\text{imp}(\b{p},\b{k})\right\rangle_\text{imp}\nn\\
&\hspace{10mm}=\frac{S(S+1)n_\text{imp}J^2}{12V}\sum_{\b{p}\alpha}\c{M}_\alpha^{\b{kp}}\c{G}^{(0)}(\b{p},i\omega_n)\c{M}_\alpha^{\b{pk}}.
\end{align}
An explicit evaluation of the sum over $\b{p}$ in Eq.~(\ref{Sigma2}) reveals that the effect of the second-order self-energy is simply to replace $\omega_n$ and $\Delta$ in the clean Green's function (\ref{Ginvclean}) by renormalized parameters $\tilde{\omega}_n$ and $\tilde{\Delta}_n$ which depend on the Matsubara frequency $\omega_n$ but not on the momentum $\b{k}$. Since the original form of the Green's function is preserved, one can thus immediately promote the second-order result to the self-consistent Born approximation, in which the self-energy has the form of Eq.~(\ref{Sigma2}) but with the internal Green's function replaced by the full disorder-averaged Green's function,
\begin{align}\label{SigmaSCBA}
\Sigma(\b{k},i\omega_n)=\frac{S(S+1)n_\text{imp}J^2}{12V}\sum_{\b{p}\alpha}\c{M}_\alpha^{\b{kp}}\c{G}(\b{p},i\omega_n)\c{M}_\alpha^{\b{pk}}.
\end{align}
In other words, we use an ansatz of the form
\begin{align}\label{GSCBA}
\c{G}(\b{k},i\omega_n)=\left(i\tilde{\omega}_n-H_\text{MF}(\b{k},\tilde{\Delta}_n)\right)^{-1},
\end{align}
where $H_\text{MF}(\b{k},\tilde{\Delta}_n)$ is the mean-field Hamiltonian matrix (\ref{HBdGmuBig}) or (\ref{HBdGmu0}), but with $\Delta$ replaced by the frequency-dependent quantity $\tilde{\Delta}_n$. Upon substituting (\ref{GSCBA}) in the Dyson equation (\ref{dyson}), and using Eq.~(\ref{SigmaSCBA}), we obtain self-consistent equations for the functions $\tilde{\omega}_n$ and $\tilde{\Delta}_n$.

\section{Proximity-induced superconductivity}
\label{sec:prox}

We will start by studying proximity-induced superconductivity. As argued before, the parameter $\Delta$ appearing in the mean-field Hamiltonians (\ref{HBdGmuBig}) and (\ref{HBdGmu0}) is a constant imposed by the bulk superconductor. It should be viewed as the maximal possible value of the pair amplitude that can be induced on the topological surface; the actual pair amplitude $\Delta_\text{eff}$ is reduced relative to $\Delta$ by disorder (see Fig.~\ref{fig:Delta_eff}). We will treat $\Delta$ as a tuning parameter, which can in principle be varied by choosing a bulk superconductor with a different gap.

\subsection{Chemical potential away from the Dirac point}
\label{sec:prox_finite_mu}

\begin{figure}[t]
\begin{center}
\includegraphics[width=0.75\linewidth]{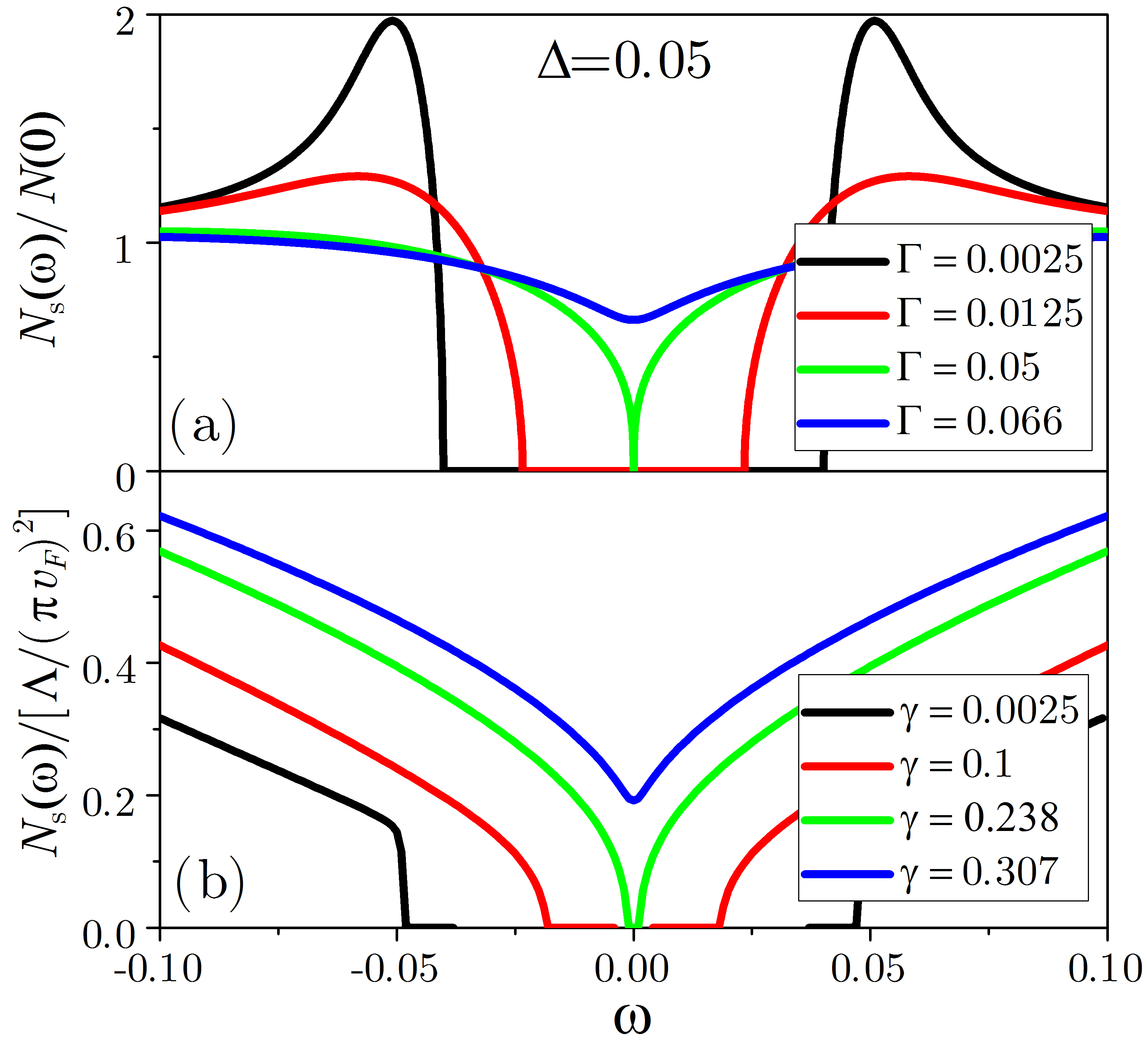}
\end{center}
\caption{Density of states for the superconducting surface state in (a) the $\mu\gg\Lambda$ regime and (b) the $\mu=0$ regime, as a function of disorder strength $\Gamma$ or $\gamma$, respectively [see Eq.~(\ref{1overtauMuBig}) and (\ref{gamma})], and for a fixed superconducting order parameter $\Delta$. The quantities $\Delta,\Gamma,\omega$ are given in units of $\Lambda$.} 
\label{dos}
\end{figure}

For $\mu\gg\Lambda$, the ansatz (\ref{GSCBA}) for the disorder-averaged Green's function becomes
\begin{align}\label{GMuBig}
\c{G}(\b{k},i\omega_n)=\frac{-i\tilde{\omega}_n-\xi_\b{k}\tau_3+\tilde{\Delta}_n\b{\tau}\cdot\hat{\b{k}}}{\tilde{\omega}_n^2+\xi_\b{k}^2+\tilde{\Delta}_n^2}.
\end{align}
Performing the sum over $\b{p}$ in Eq.~(\ref{SigmaSCBA}), the self-energy in the self-consistent Born approximation is given by
\begin{eqnarray}\label{SelfMuBig}
\Sigma(\b{k},i\omega_n)
&=&\frac{\Gamma}{2}\frac{-i\tilde{\omega}_n+\tilde{\Delta}_n\b{\tau}\cdot\hat{\b{k}}}{\sqrt{\tilde{\omega}_n^2+\tilde{\Delta}_n^2}},
\end{eqnarray}
where
\begin{align}\label{1overtauMuBig}
\Gamma=S(S+1)\pi n_\text{imp}J^2N(0)
\end{align}
is the impurity scattering rate, with units of inverse time. Substituting (\ref{GMuBig}) and (\ref{SelfMuBig}) in the Dyson equation (\ref{dyson}), we obtain the self-consistency conditions
\begin{figure}[t]
\begin{center}
\includegraphics[width=0.75\linewidth]{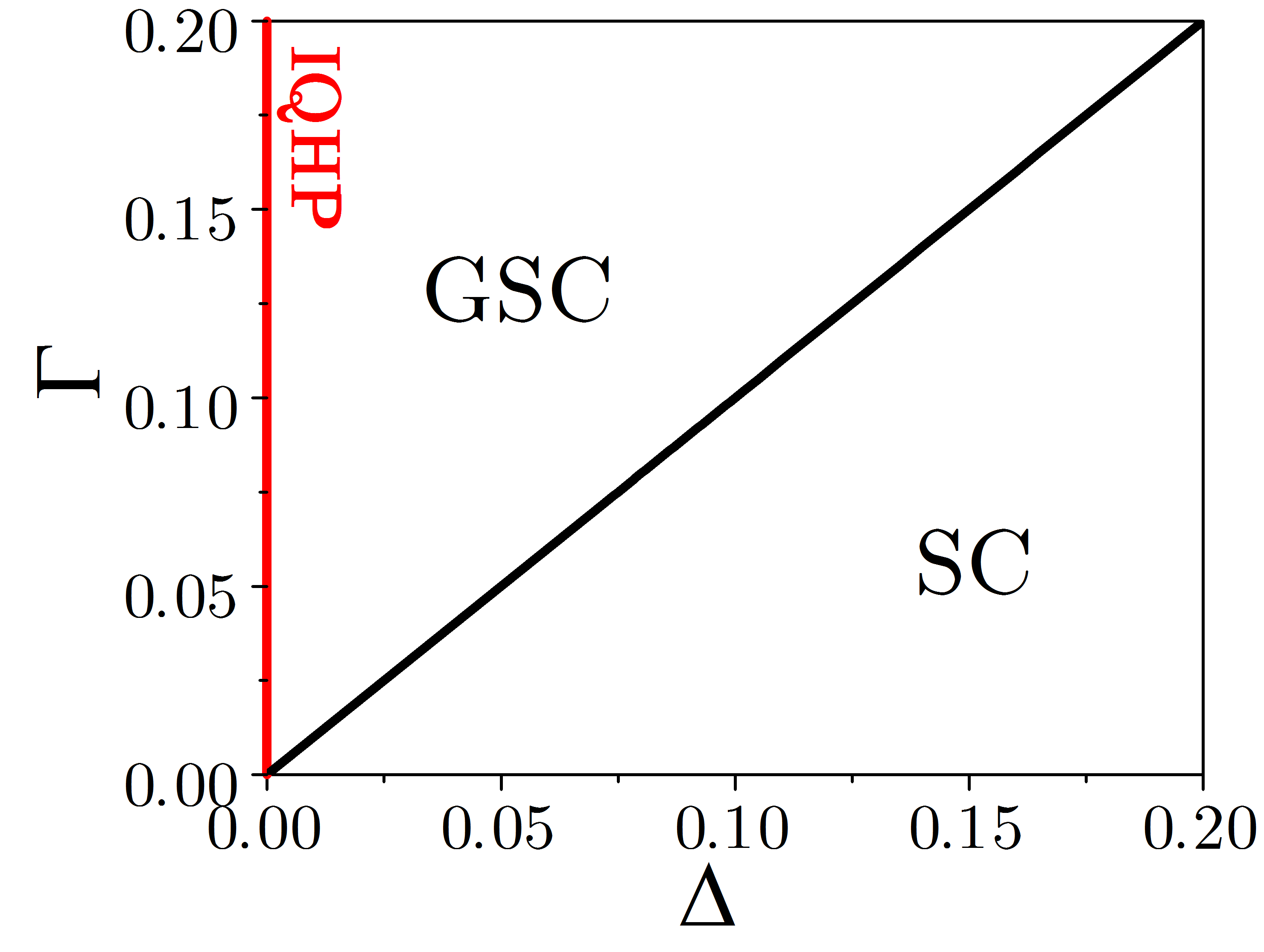}
\end{center}
\caption{Phase diagram of the surface state in the $\mu\gg\Lambda$ regime with respect to the maximal proximity-induced superconducting order parameter $\Delta$ and impurity scattering rate $\Gamma$ in units of $\Lambda$. A phase transition from a gapped helical superconductor (SC) to a gapless helical superconductor (GSC) is observed for $\Gamma>\Delta$. In the absence of superconductivity ($\Delta=0$), the disordered surface state is in the universality class of the integer quantum Hall plateau transition (IQHP).} 
\label{tDprox1}
\end{figure}
\begin{eqnarray}
    \tilde{\omega}_n&=&\omega_n+\frac{\Gamma}{2}\frac{\tilde{\omega}_n}{\sqrt{\tilde{\omega}_n^2+\tilde{\Delta}_n^2}},\label{scmg01}\\
    \tilde{\Delta}_n&=&\Delta-\frac{\Gamma}{2}\frac{\tilde{\Delta}_n}{\sqrt{\tilde{\omega}_n^2+\tilde{\Delta}_n^2}},\label{scmg02}
    \end{eqnarray}
which have precisely the same form as for a conventional $s$-wave superconductor doped with magnetic impurities~\cite{skalski1964}. Defining two new parameters $\tilde{u}_n=\tilde{\omega}_n/\tilde{\Delta}_n$ and $u_n=\omega_n/\Delta$, \eq{scmg01} and (\ref{scmg02}) combine into a single self-consistency equation
\begin{align}\label{SelfConsistentMuBig}
\tilde{u}_n=u_n+\frac{\Gamma}{\Delta}\frac{\tilde{u}_n}{\sqrt{1+\tilde{u}_n^2}}.
\end{align}
The density of states $N_s(\omega)$ of the superconducting surface state can be determined from the electron component of the Nambu Green's function (\ref{GMuBig}) analytically continued to real frequencies,
\begin{align}
N_s(\omega)=-\frac{1}{\pi}\im\sum_\b{k}\c{G}_{11}(\b{k},\omega+i\delta),
\end{align}
where $\delta$ is a positive infinitesimal. Using the self-consistency condition (\ref{SelfConsistentMuBig}), we obtain
\begin{align}\label{DOSeq}
\frac{N_s(\omega)}{N(0)}=\frac{\Delta}{\Gamma}\im\left(\left.i\tilde{u}_n\right|_{i\omega_n\rightarrow\omega+i\delta}\right),
\end{align}
where $N(0)$ is the density of states at the Fermi level in the normal (i.e., metallic) state.

\begin{figure}[t]
\begin{center}
\includegraphics[width=0.75\linewidth]{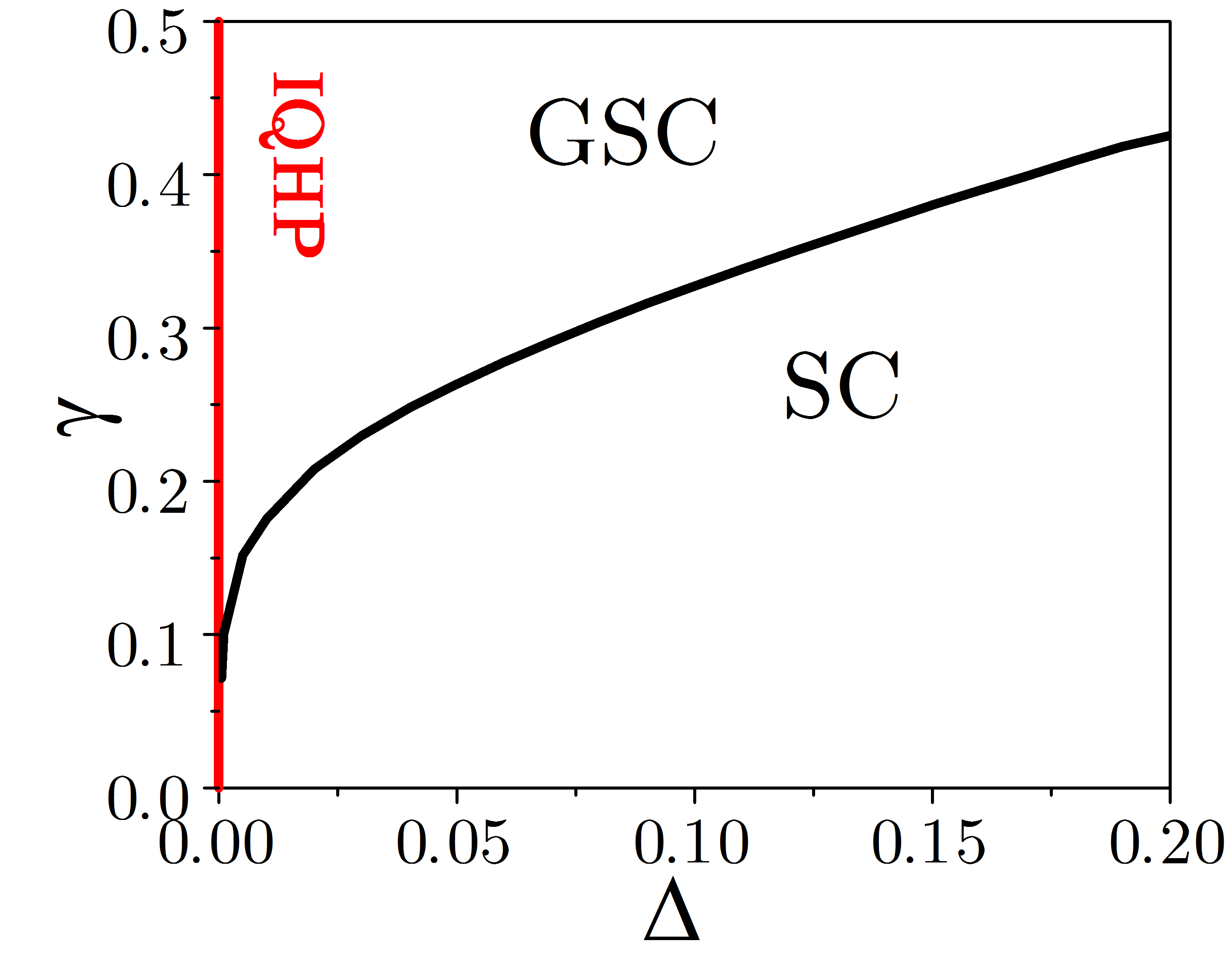}
\end{center}
\caption{Phase diagram of the surface state in the $\mu=0$ regime. Here $\gamma$ is a dimensionless measure of the disorder strength; notations are otherwise the same as in Fig.~(\ref{tDprox1}).} 
\label{tDprox2}
\end{figure}

In a clean superconductor, the superconducting gap is dictated by $\Delta$, but in a dirty superconductor this one-to-one correspondence can break down~\cite{abrikosov1961}. In Fig.~\ref{dos}(a) we plot the density of states of the superconducting surface in the $\mu\gg\Lambda$ regime, for various disorder strengths and a fixed value of $\Delta/\Lambda=0.05$~\footnote{Strictly speaking, in the $\mu\gg\Lambda$ regime the cutoff $\Lambda$ does not appear explicitly in the self-consistency condition (\ref{SelfConsistentMuBig}) and the density of states is a universal function of $\omega/\Delta$ and $\Gamma/\Delta$. However, for $\mu=0$ the cutoff appears explicitly in the self-consistency conditions, and the density of states depends on $\omega/\Lambda$ and $\Delta/\Lambda$ separately. For the purpose of plotting both curves on the same graph we thus measure all energies in units of $\Lambda$.}. For the smallest scattering rate, corresponding to weak disorder, a hard gap of magnitude $\approx 2\Delta$ is flanked by well-formed BCS coherence peaks. As the scattering rate increases, first the coherence peaks smear out while the gap is reduced below $2\Delta$. Eventually, at a critical value of the scattering rate $\Gamma_c=\Delta$ the gap closes. For stronger disorder still, i.e., $\Gamma>\Delta$, the density of states at the Fermi level becomes nonzero and is given by $N_s(0)/N(0)=\sqrt{1-(\Delta/\Gamma)^2}$, although $\Delta$ is still non-vanishing. One thus obtains a gapless helical superconductor for $\Gamma>\Delta$, giving rise to the phase diagram shown in \Fig{tDprox1}.

For $\Delta=0$, we have a non-superconducting topological surface with quenched magnetic disorder, which can be mapped to a single flavor of 2D Dirac fermions with a random gauge field and a random mass~\cite{liu2009,foster2012}. Although we have not explicitly included a random scalar potential (i.e., a random chemical potential) in our impurity Hamiltonian (\ref{Himp}), in the presence of a random gauge field and random mass this type of disorder will be generated by renormalization~\cite{foster2012,ludwig1994,ostrovsky2006}. In the presence of all three types of disorder, one is in class A (unitary class) of the Altland-Zirnbauer classification~\cite{altland1997}. The system is believed to flow at low energies to the critical point of the integer quantum Hall plateau transition (IQHP)~\cite{ludwig1994,nomura2008}, which is characterized by a nonzero density of states~\cite{huckestein1995}. This conclusion is expected to hold for both the doped surface ($\mu\gg\Lambda$) and for a chemical potential at the Dirac point ($\mu=0$).

\subsection{Chemical potential at the Dirac point}
\label{sec:prox_mu_zero}

For $\mu=0$, we must retain the full $4\times 4$ structure of the Nambu propagators with two helicities, and the disorder-averaged Green's function (\ref{GSCBA}) in the self-consistent Born approximation is given by
\begin{align}\label{GexplicitMu0}
\c{G}(\b{k},i\omega_n)=\frac{-i\tilde{\omega}_n-\epsilon_\b{k}\sigma_3\otimes\tau_3+\tilde{\Delta}_n\sigma_3\otimes\b{\tau}\cdot\hat{\b{k}}}
{\tilde{\omega}_n^2+\epsilon_\b{k}^2+\tilde{\Delta}_n^2}.
\end{align}
Unlike the doped case $\mu\gg\Lambda$, because here the normal-state density of states $N(\epsilon)=|\epsilon|/2\pi v_F^2$ vanishes at the Fermi level we must retain its energy dependence when performing the sum over $\b{p}$ in Eq.~(\ref{SigmaSCBA}). The self-energy becomes
\begin{eqnarray}\label{SigmaExplicitMu0}
\Sigma(\b{k},i\omega_n)&=&\frac{\gamma}{2}\ln\left(1+\frac{\Lambda^2}{\tilde{\omega}_n^2+\tilde{\Delta}_n^2}\right)\nn\\
&&\times\left(-i\tilde{\omega}_n+\tilde{\Delta}_n\sigma_3\otimes\b{\tau}\cdot\hat{\b{k}}\right),
\end{eqnarray}
where the parameter
\begin{align}\label{gamma}
\gamma=\frac{S(S+1)n_\text{imp}J^2}{\pi v_F^2}
\end{align}
is similar to the impurity scattering rate (\ref{1overtauMuBig}), but is a \emph{dimensionless} measure of the disorder strength appropriate for 2D Dirac systems with a vanishing density of states at the Fermi level~\cite{shon1998}. Inserting Eq.~(\ref{GexplicitMu0}) and (\ref{SigmaExplicitMu0}) in the Dyson equation (\ref{dyson}), we obtain the self-consistency conditions
 \begin{eqnarray}
\tilde{\omega}_n&=&\omega_n+\frac{\gamma\tilde{\omega}_n}{2}\ln\left(1+\frac{\Lambda^2}{\tilde{\omega}_n^2+\tilde{\Delta}_n^2}\right),\label{scm0Omg}\\
\tilde{\Delta}_n&=&\Delta-\frac{\gamma\tilde{\Delta}_n}{2}\ln\left(1+\frac{\Lambda^2}{\tilde{\omega}_n^2+\tilde{\Delta}_n^2}\right).\label{scm0Del}
\end{eqnarray}
The density of states in the superconducting state is
\begin{align}
N_s(\omega)&=-\frac{1}{\pi}\sum_\b{k}\left[\c{G}_{11}(\b{k},\omega+i\delta)+\c{G}_{33}(\b{k},\omega+i\delta)\right]\nn\\
&=\frac{1}{(\pi v_F)^2}\im\left(\left.i\tilde{\omega}_n X_n\right|_{i\omega_n\rightarrow\omega+i\delta}\right),
\end{align}
where we define
\begin{align}\label{xn1}
X_n\equiv \ln\left(1+\frac{\Lambda^2}{\tilde{\omega}_n^2+\tilde{\Delta}_n^2}\right).
\end{align}
Expressing Eq.~(\ref{scm0Omg})-(\ref{scm0Del}) as a function of $X_n$,
\begin{align}\label{xnwd}
\tilde{\Delta}_n=\frac{\Delta}{1+\gamma X_n/2},\hspace{5mm}
\tilde{\omega}_n=\frac{\omega_n}{1-\gamma X_n/2}, 
\end{align}
and substituting \eq{xnwd} back into \eq{xn1}, we combine the two self-consistent equations (\ref{scm0Omg})-(\ref{scm0Del}) into a single equation for the parameter $X_n$. For a given value of $\Delta$ and $\gamma$ one can obtain $X_n$ for each (analytically continued) Matsubara frequency $i\omega_n=\omega+i\delta$ and calculate the density of states from
\begin{equation}
N_s(\omega)=\frac{1}{(\pi v_F)^2}\im\left[(\omega+i\delta)\left.\frac{X_n}{1-\gamma X_n/2}\right|_{i\omega_n\rightarrow\omega+i\delta}\right].
\end{equation}

In \Fig{dos}(b) we plot the density of states (in units of $\Lambda/(\pi v_F)^2$) of the superconducting surface with chemical potential at the Dirac point, again for a fixed value of $\Delta/\Lambda=0.05$ and for various values of the dimensionless disorder strength $\gamma$. For $\mu=0$, there are no BCS coherence peaks even for very small $\gamma$ due to the vanishing density of states at the Fermi level in the clean semimetal, but there is still a hard gap of magnitude $\approx 2\Delta$. As the disorder strength increases, the gap decreases until it closes at the Fermi level at a critical value of the disorder strength. Beyond this critical value one has a gapless helical superconductor, as shown in the phase diagram of Fig.~\ref{tDprox2}. By contrast with the $\mu\gg\Lambda$ case, here the critical disorder strength depends nonlinearly on the superconducting order parameter.
 
\begin{figure}[t]
\begin{center}
\includegraphics[width=0.85\columnwidth]{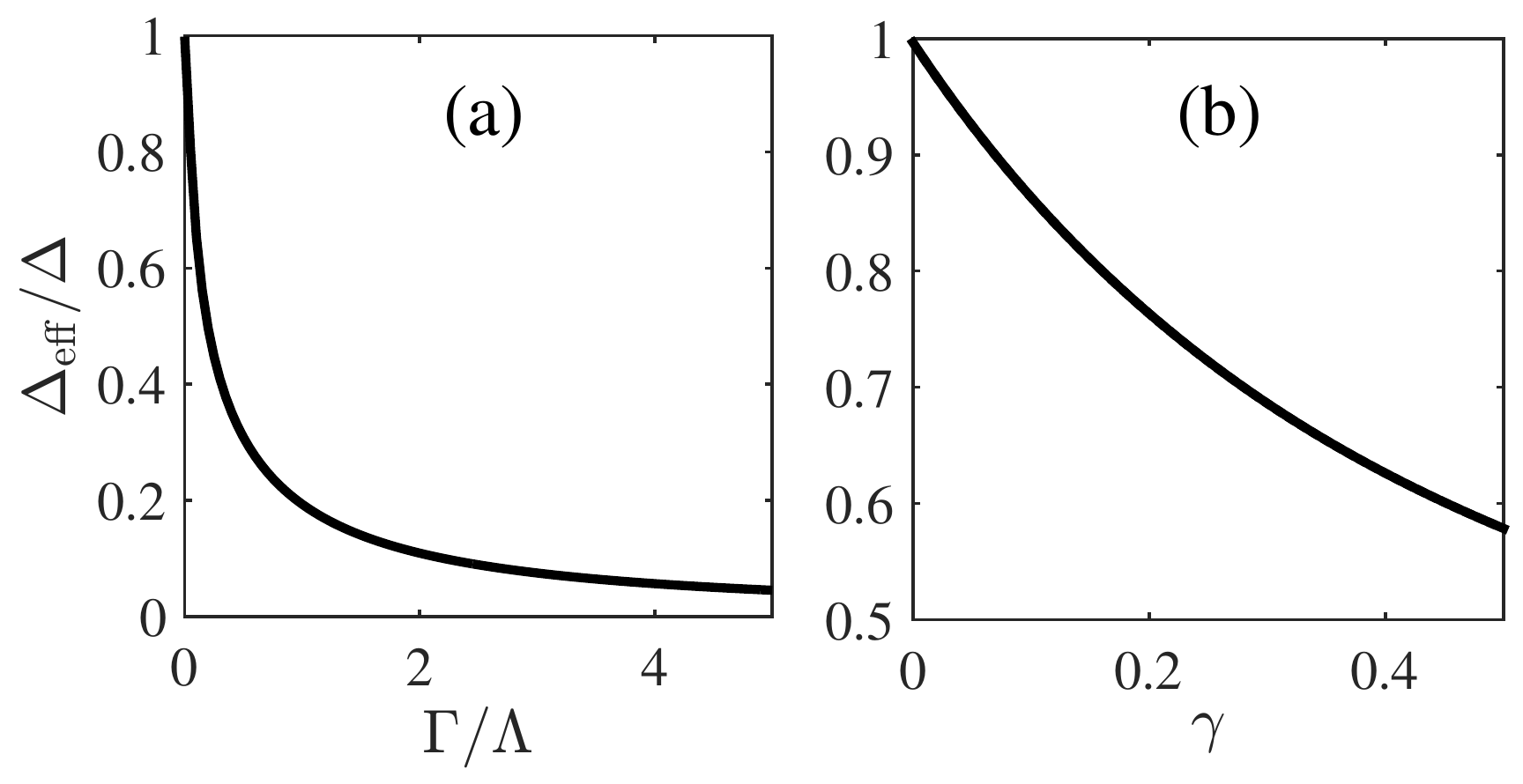}
\end{center}
\caption{In the presence of disorder, the true pair amplitude $\Delta_\text{eff}$ is reduced relative to the induced pair amplitude $\Delta$ in the clean limit, for both (a) large doping and (b) zero doping. Plots are shown for $\Delta/\Lambda=0.05$ and zero temperature.}
\label{fig:Delta_eff}
\end{figure}

Figures~\ref{tDprox1} and \ref{tDprox2} suggest that some form of proximity-induced superconductivity, either gapped or gapless, persists on the topological insulator surface regardless of the strength of disorder. However, when interpreting these phase diagrams two caveats are in order. First, our study is perturbative in the disorder strength; for sufficiently strong disorder localization effects may become important and our conclusions may not hold. Second, in these phase diagrams $\Delta$ should be interpreted as an external tuning parameter rather than an actual measure of the superconducting correlations in the system. Indeed, the true (impurity-averaged) pair amplitude $\Delta_\text{eff}\propto\langle\psi_\uparrow(\b{r})\psi_\downarrow(\b{r})\rangle$ is gradually weakened with increasing disorder strength relative to the clean-limit value $\Delta$ (see Fig.~\ref{fig:Delta_eff} and Appendix~\ref{app}). Thus for sufficiently large disorder the system may still have nonzero superconducting correlations, but the latter may be too small to be observable.

\section{Intrinsic superconductivity}
\label{sec:intrinsic}

\begin{figure}[t]
\begin{center}
\includegraphics[width=0.75\linewidth]{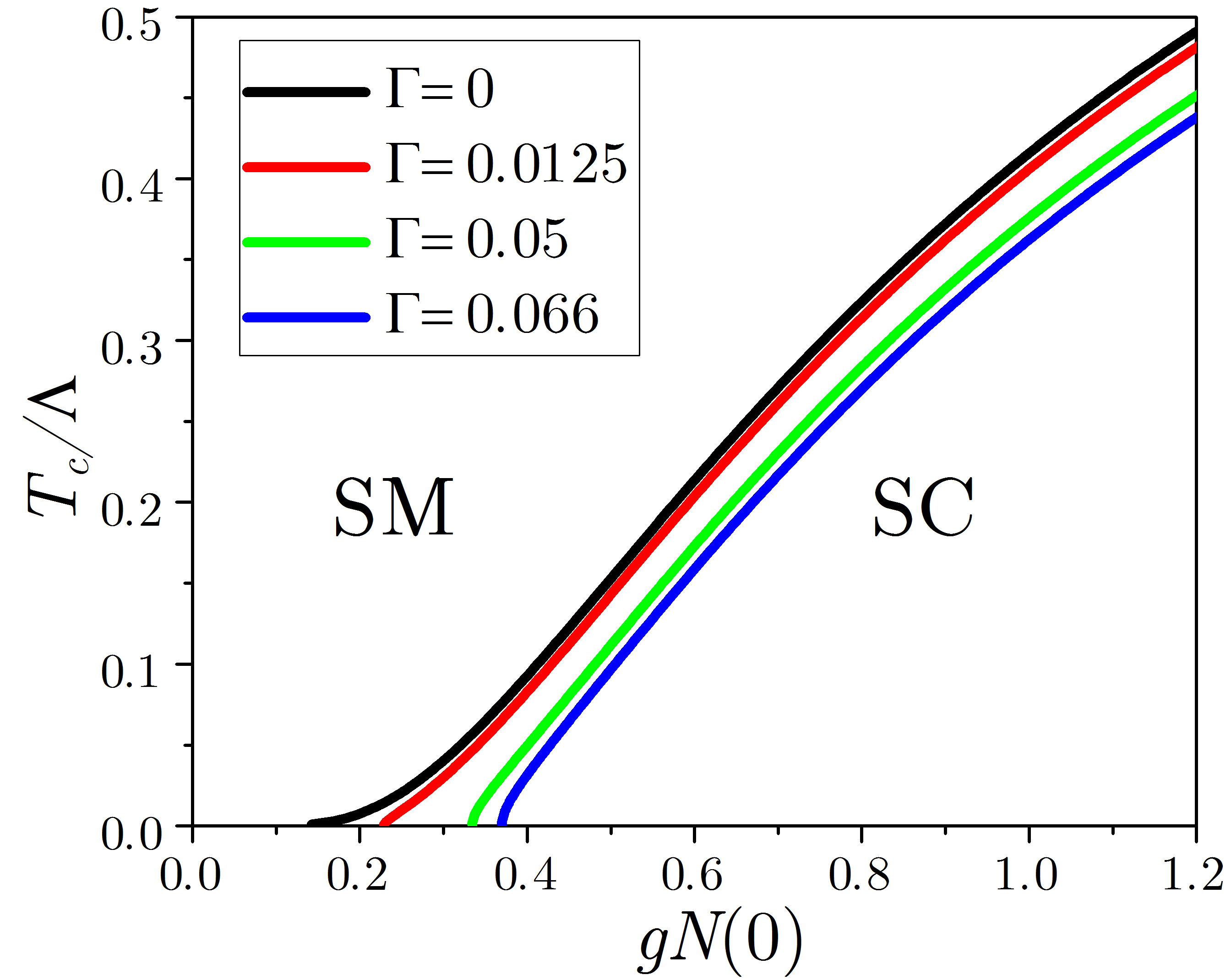}
\end{center}
\caption{Superconducting transition temperature $T_c$ of the disordered system as a function of interaction strength $g$ in the $\mu\gg\Lambda$ regime, for various values of the impurity scattering rate $\Gamma$. Here no distinction is made between the gapless and gapped helical superconductor regions.} 
\label{gtdirtymg0}
\end{figure}

In the previous section we treated the pair amplitude $\Delta$ appearing in the Hamiltonian as a constant imposed by the bulk superconductor. This parameter could be interpreted as the induced pair amplitude in the clean limit, while the actual pair amplitude $\Delta_\text{eff}$ was somewhat reduced by disorder relative to $\Delta$ (Fig.~\ref{fig:Delta_eff}). However, if we are dealing with an intrinsic superconductor, $\Delta$ is not an externally imposed parameter but depends on the temperature $T$ and the interaction strength $g$. The main question thus becomes how magnetic impurities modify the superconducting transition temperature $T_c$ from its value in the clean limit [see Eq.~(\ref{mg0cleangt}) and (\ref{Tc0DiracPt})] and, in the case of a Fermi level at the Dirac point, the critical interaction strength (\ref{gc0}). To address these questions, one needs to supplement the self-consistency conditions (\ref{SelfConsistentMuBig}) in the $\mu\gg\Lambda$ limit or (\ref{scm0Omg})-(\ref{scm0Del}) in the $\mu=0$ limit with a self-consistency equation for the superconducting order parameter itself.

\subsection{Chemical potential away from the Dirac point}

In the $\mu\gg\Lambda$ limit, the self-consistency equation for the order parameter $\Delta$ is obtained by evaluating the right-hand side of Eq.~(\ref{GapEqlargemu}) with the disorder-averaged Green's function (\ref{GMuBig}),
\begin{align}\label{GapEqDisMuBig}
\Delta=\pi gTN(0)\sum_{i\omega_n}\frac{1}{\sqrt{1+\tilde{u}_n^2}}.
\end{align}
In conjunction with Eq.~(\ref{SelfConsistentMuBig}), this equation can be used to determine the superconducting transition temperature $T_c$ in the disordered system, defined as the temperature above which the order parameter $\Delta$ vanishes. Solving Eq.~(\ref{GapEqDisMuBig}) and (\ref{SelfConsistentMuBig}) simultaneously in the limit $\Delta\rightarrow 0$, we obtain the relation
\begin{equation}\label{AGTc}
    \ln\left(\frac{T_c^0}{T_c}\right)=\psi\left(\frac{1}{2}+\frac{\Gamma}{2\pi T_c}\right)-\psi\left(\frac{1}{2}\right),
\end{equation}
where $\psi(z)$ is the digamma function, which is precisely the Abrikosov-Gor'kov result for a conventional $s$-wave superconductor with magnetic impurities~\cite{abrikosov1961}. In the limit of $\Gamma\rightarrow 0$ we recover \eq{mg0cleangt}. The right-hand side is always positive, indicating that disorder always leads to a reduction in the transition temperature $T_c$. For weak disorder $\Gamma\ll T_c^0$, Eq.~(\ref{AGTc}) predicts a linear reduction in $T_c$, which can be expressed in the form
\begin{align}\label{AGrate}
\frac{dT_c}{d\Gamma}=-\frac{\pi}{4},
\end{align}
i.e., the suppression rate is universal, which has been observed experimentally in conventional superconductors~\cite{woolf1965}.

\begin{figure}[t]
\begin{center}
\includegraphics[width=0.70\linewidth]{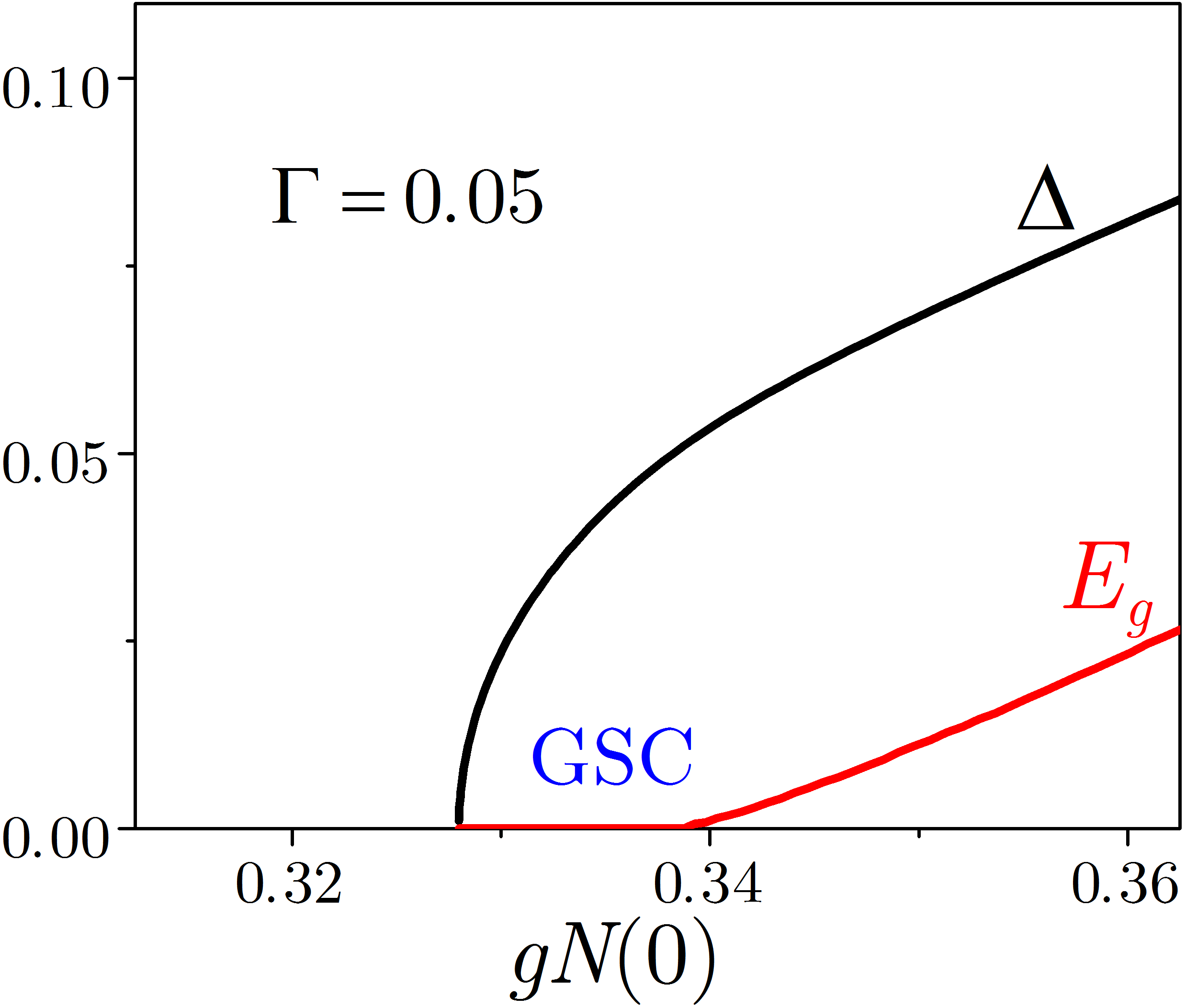}
\end{center}
\caption{Superconducting order parameter $\Delta$ and quasiparticle gap $E_g$ as a function of interaction strength $g$ near zero temperature, in the $\mu\gg\Lambda$ regime and for an impurity scattering rate $\Gamma=0.05$. All energies are in units of $\Lambda$.} 
\label{gscmg0}
\end{figure}

For generic values of the disorder strength, the critical temperature can be determined from Eq.~(\ref{AGTc}) numerically (Fig.~\ref{gtdirtymg0}). With increasing disorder, the critical temperature is reduced at a fixed interaction strength $g$, and the critical interaction strength $g$ is increased at a fixed temperature. In particular, at zero temperature and in the presence of disorder there exists a finite critical interaction strength $g_c$ one must exceed to develop superconductivity. In the weak disorder limit, the value of $g_c$ can be determined from Eq.~(\ref{AGrate}),
\begin{align}
g_cN(0)=\frac{1}{\ln\left(\frac{8e^\gamma}{\pi^2}\frac{\Lambda}{\Gamma}\right)}.
\end{align}
In this expression $\gamma$ denotes Euler's constant, not to be confused with the dimensionless disorder strength in Eq.~(\ref{gamma}).

In Fig.~\ref{gtdirtymg0}, we did not make a distinction between gapped and gapless superconducting regions. To determine what type of superconductor develops below $T_c$, we can use Eq.~(\ref{DOSeq}) to calculate the density of states of the superconductor, based on the value of the superconducting order parameter $\Delta$ found from the solution of Eq.~(\ref{GapEqDisMuBig}). In \Fig{gscmg0} we plot $\Delta$ as well as the quasiparticle gap $E_g$ determined from the electronic density of states, as a function of the BCS interaction strength $g$. One clearly sees the gapless superconducting region (GSC) where $E_g=0$ while $\Delta\neq 0$. Since the interaction strength only indirectly affects the density of states through $\Delta$, the phase diagram in the $\Gamma$-$\Delta$ plane and the density of states in the superconducting phases for the same values of $\Gamma$ and $\Delta$ are the same as in Sec.~\ref{sec:prox_finite_mu}.

\subsection{Chemical potential at the Dirac point}

\begin{figure}[t]
\begin{center}
\includegraphics[width=0.75\linewidth]{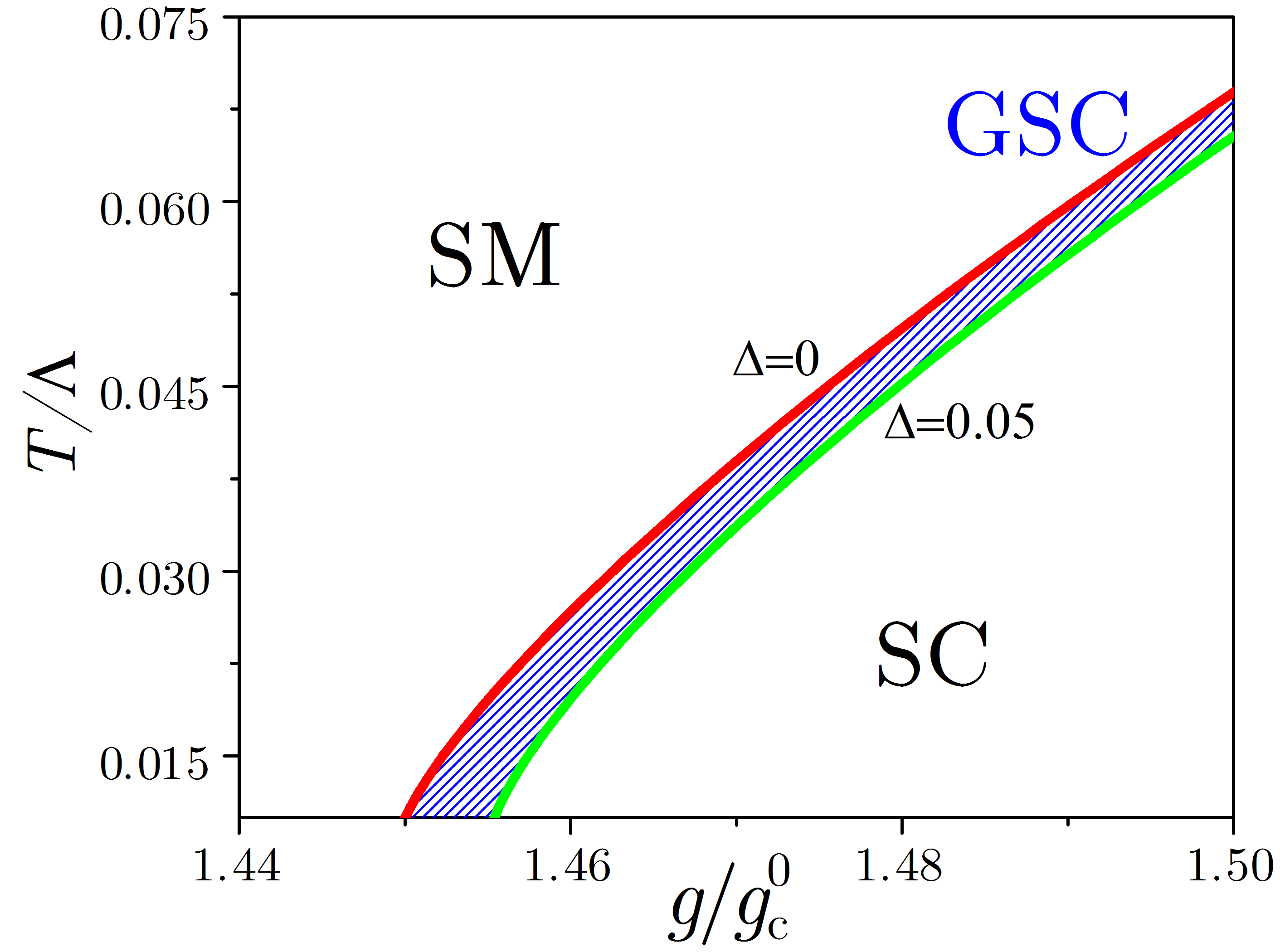}
\end{center}
\caption{Phase diagram with respect to interaction strength $g$ and temperature $T$, in the $\mu=0$ regime and for $\gamma=0.25$, showing the semimetallic (SM), gapless (GSC, blue hatched region) and gapped (SC) helical superconductor regions. The value of the superconducting order parameter $\Delta$ at the GSC-SC phase transition is 0.05. All energies are in units of $\Lambda$.} 
\label{phasem0}
\end{figure}

\begin{figure}[t]
\begin{center}
\includegraphics[width=0.75\linewidth]{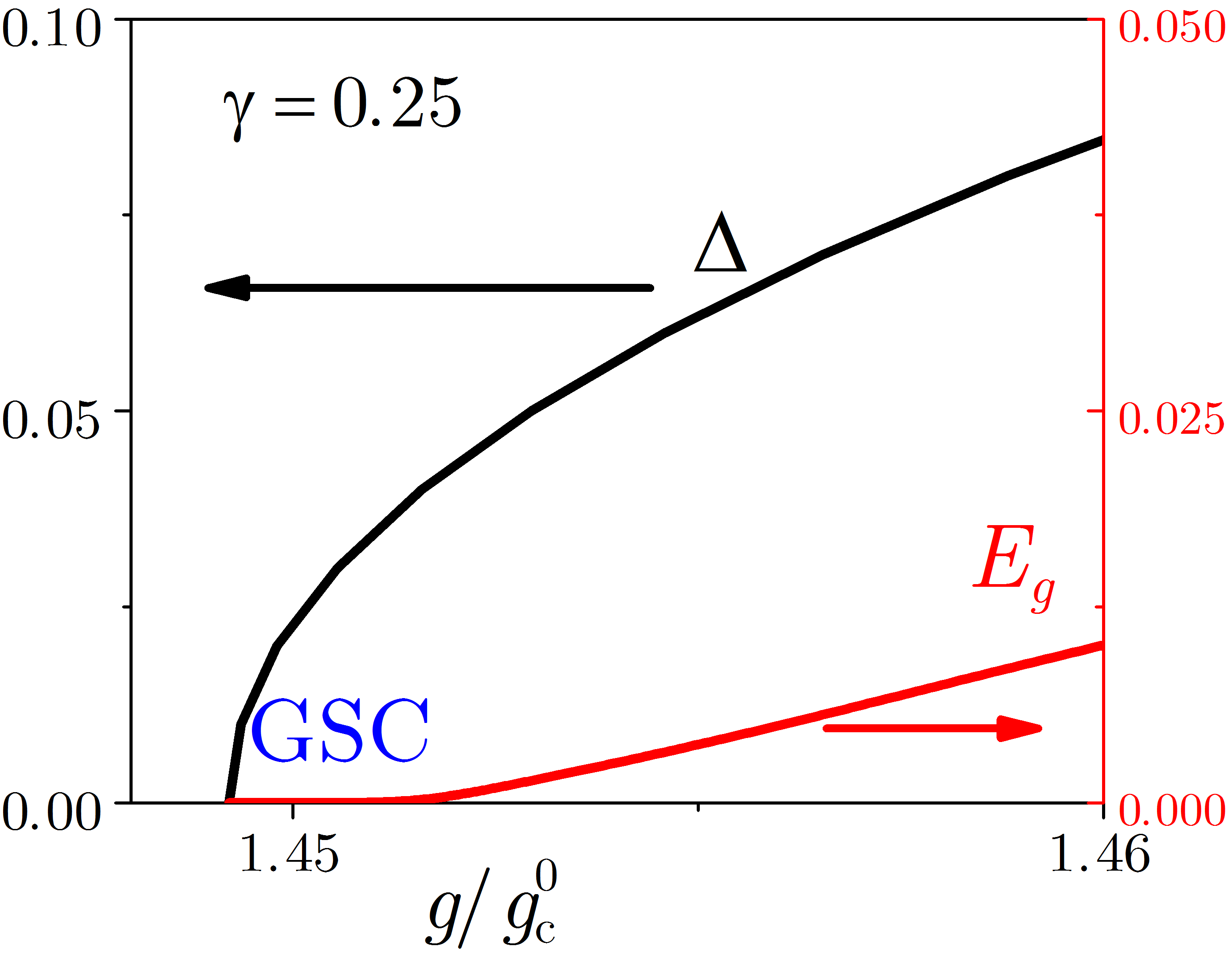}
\end{center}
\caption{Superconducting order parameter $\Delta$ and quasiparticle gap $E_g$ as a function of interaction strength $g$ near zero temperature, in the $\mu=0$ regime and for disorder strength $\gamma=0.25$. All energies are in units of $\Lambda$.} 
\label{gaplessSC}
\end{figure}

In the $\mu=0$ limit, we evaluate the right-hand side of Eq.~(\ref{gapeq}) with the disordered-averaged Green's function (\ref{GexplicitMu0}), to obtain
\begin{equation}\label{GapEqIntrMu0}
    \Delta=\frac{g}{g_c^0}\frac{T}{\Lambda}\sum_{i\omega_n}\tilde{\Delta}_n\ln\left(1+\frac{\Lambda^2}{\tilde{\omega}_n^2+\tilde{\Delta}_n^2}\right), 
\end{equation}
where $g_c^0$ is the critical interaction strength at zero temperature, defined in Eq.~(\ref{gc0}). As in the previous section, to determine the phase diagram of the disordered superconductor we must solve Eq.~(\ref{GapEqIntrMu0}) together with the self-consistency conditions (\ref{scm0Omg})-(\ref{scm0Del}). In Sec.~\ref{sec:prox_mu_zero} we showed these two equations can be combined into a single equation for the parameter $X_n$ defined in Eq.~(\ref{xn1}). Using Eq.~(\ref{xnwd}), the self-consistent equation (\ref{GapEqIntrMu0}) for the superconducting order parameter can be expressed in terms of $X_n$ alone as
\begin{equation} 
\frac{g_c^0}{g}=\frac{T}{\Lambda}\sum_{i\omega_n}\frac{X_n}{1+\gamma X_n/2}. 
\label{geq}
\end{equation}
The phase diagram for $\gamma=0.25$ in the $g$-$T$ plane, shown in \Fig{phasem0}, depicts the $\Delta=0$ line in red separating the semimetallic (SM) and helical superconducting phases. In addition, the green line, corresponding to $\Delta=0.05$, separates the helical superconducting phase into gapped (SC) and gapless (GSC) regions. With increasing disorder strength (not shown), the transition temperature is lowered for a fixed interaction strength $g$, and the critical interaction increases for a fixed temperature. In particular, pairing at zero temperature remains a quantum critical phenomenon in the disordered system, but the quantum critical interaction strength $g_c$ is larger than that [Eq.~(\ref{gc0})] in the clean system. In Fig.~\ref{gaplessSC} we plot the superconducting order parameter $\Delta$ and the quasiparticle gap $E_g$ obtained from the density of states. Here again, one clearly sees the gapless superconducting region (GSC) where $E_g=0$ but $\Delta\neq 0$. As for $\mu\gg\Lambda$, the form of the density of states $N_s(\omega)$ for the intrinsic superconductor for any given values of the disorder strength $\gamma$ and the order parameter $\Delta$, as well as the phase diagram in the $\gamma$-$\Delta$ plane, are the same as for the proximity-induced superconductor (Sec.~\ref{sec:prox_mu_zero}).

What is the $\mu=0$ analog of the universal Abrikosov-Gor'kov $T_c$ suppression rate (\ref{AGrate})? The first difference between the $\mu\gg\Lambda$ and $\mu=0$ regimes is that in the clean limit, one has the infinitesimal Cooper instability for $\mu\gg\Lambda$ while pairing requires overcoming a threshold interaction strength for $\mu=0$. A first question one may thus ask is whether the current theory predicts a universal rate of increase of the zero-temperature critical BCS coupling (\ref{gc0}) with disorder strength. Solving Eq.~(\ref{GapEqIntrMu0}) at $T=0$, and for weak disorder, we find the universal rate
\begin{align}\label{DiracRategc}
\frac{d\ln g_c}{d\gamma}=3\ln 2,
\end{align}
which can be interpreted as a quantum critical version of (\ref{AGrate}). The second difference between the two regimes of chemical potential is that in the $\mu\gg\Lambda$ regime the impurity scattering rate $\Gamma$ and the superconducting transition temperature $T_c$ have the same units of energy (we are working in units in which $\hbar=k_B=1$), thus $dT_c/d\Gamma$ is dimensionless and can be universal, while here the disorder strength $\gamma$ is dimensionless and thus $dT_c/d\gamma$ is not. This is directly related to the different scaling of the normal-state density of states with energy near the Fermi level in both regimes. Although the ratio $dT_c/d\gamma$ cannot be universal, perhaps $d\ln T_c/d\gamma$ can.

In order to derive an analytical expression for the suppression of $T_c$ at zero chemical potential, we first invoke the following simple argument. Assume that in the disordered system $T_c$ vanishes linearly near $g_c$, as it does in the clean system [Eq.~(\ref{Tc0DiracPt})],
\begin{align}
\frac{T_c}{\Lambda}=\alpha\left(\frac{g-g_c}{g_c}\right),\hspace{5mm}
\frac{g-g_c}{g_c}\ll 1,
\end{align}
where we consider $g>g_c$ and $\alpha$ is a constant that depends on the disorder strength $\gamma$. This linear vanishing is expected from mean-field theory and is consistent with our numerical observations (see, e.g., Fig.~\ref{phasem0}). Considering an infinitesimal amount of disorder $d\gamma$, we can write $T_c=T_c^0+dT_c$, $g_c=g_c^0+dg_c$, and $\alpha=\alpha_0+\alpha_0'd\gamma$, where $dg_c=3\ln 2\cdot g_c^0d\gamma$ from Eq.~(\ref{DiracRategc}) and $\alpha_0,\alpha_0'$ are constants to be determined. Keeping only terms to leading order in $d\gamma$ and $(g-g_c)/g_c$, we obtain
$d\ln T_c/d\gamma=-3\ln 2\cdot g_c/(g-g_c)+\alpha_0'\cdot 2\ln 2$. To determine the unknown constant $\alpha_0'$, we expand Eq.~(\ref{GapEqIntrMu0}) to leading order in $\gamma$ and $(g-g_c)/g_c$ and find precisely the same result, but with $\alpha_0'=0$. We thus obtain
\begin{align}\label{DiracRateTc}
\frac{d\ln T_c}{d\gamma}=-3\ln 2\left(\frac{g-g_c}{g_c}\right)^{-1},
\hspace{5mm}\frac{g-g_c}{g_c}\ll 1,
\end{align}
which can be thought of as the Dirac analog of the Abrikosov-Gor'kov formula (\ref{AGrate}). Just like the latter, this formula is universal in the sense that it does not depend on material-specific parameters such as the Fermi velocity $v_F$ and the pairing scale $\Lambda$. It depends, however, on the dimensionless parameter $(g-g_c)/g_c$ specifying the distance to the quantum critical point. Of course, very close to the latter we expect fluctuation corrections to Eq.~(\ref{DiracRateTc}), since the linear dependence of $T_c$ on $g-g_c$ is a mean-field result.

\section{Conclusion}
\label{sec:conclusion}

In this paper we have studied the effect of random magnetic impurities on helical superconductivity in the Dirac surface states of 3D topological insulators. Such impurities disturb both the topological protection of the surface states and the formation of Cooper pairs by virtue of breaking time-reversal symmetry; thus one would doubly expect that they lead to the complete suppression of helical superconductivity. Contrary to this expectation, however, we find there exist parameter regimes in which helical superconductivity survives in the form of a gapless helical superconductor. In this gapless helical superconductor, the usual correspondence between superconducting order parameter and superconducting gap breaks down: the gap closes and the usual BCS coherence peaks disappear, but the order parameter (and thus the existence of Cooper pairs) remains. Our results indicate that in the presence of magnetic impurities, one cannot unambiguously infer the presence or absence of helical superconductivity in the topological surface states from the existence of a gap in the electronic density of states or lack thereof, as is well-known for conventional bulk superconductors~\cite{abrikosov1961}. It is also known that conventional gapless superconductors exhibit a Meissner effect, and thus can support persistent currents~\cite{skalski1964}; we speculate that the same is true of the gapless helical superconductor, and suggest attempting to perform measurements of the latter in Mn-doped Bi$_2$Se$_3$ films grown on NbSe$_2$ (or indeed, in any other superconducting topological insulator doped with magnetic impurities) to test some of the ideas put forward here.

\acknowledgements

We thank A. Chandran for a useful discussion. I.O. was supported by Alberta Innovates-Technology Futures (AITF). J.H. was supported by the China Scholarship Council (CSC). J.M. was supported by NSERC grant \#RGPIN-2014-4608, the Canada Research Chair Program (CRC), the Canadian Institute for Advanced Research (CIFAR), and the University of Alberta.

\appendix
\section{Reduction of pair amplitude due to disorder}
\label{app}

In this Appendix we explain how Fig.~\ref{fig:Delta_eff} was obtained. For proximity-induced superconductivity, the parameter $\Delta$ appearing in the mean-field Hamiltonians (\ref{HBdGmuBig}) and (\ref{HBdGmu0}) is a constant imposed by the bulk superconductor, independent of the disorder strength. However, the actual superconducting correlations on the topological insulator surface are affected by disorder. We consider the impurity-averaged local pair amplitude $\Delta_\text{eff}\propto\langle\psi_\uparrow(\b{r})\psi_\downarrow(\b{r})\rangle$ in the presence of disorder. This is determined by the anomalous Green's function $\mathcal{F}$ and thus depends on the renormalized quantity $\tilde{\Delta}_n$. Focussing on the limit of zero temperature, where the Matsubara frequency $\omega_n$ becomes continuous (we then call it $\omega$), we have
\begin{align}
\Delta_\text{eff}&\propto\int\frac{d\omega}{2\pi}\sum_\b{k}\mathcal{F}(\b{k},i\omega)\nn\\
&\propto\int\frac{d\omega}{2\pi}\int_{-\Lambda}^\Lambda d\xi\,N(\xi)\frac{\tilde{\Delta}(\omega)}{\xi^2+\tilde{\Omega}^2(\omega)+\tilde{\Delta}^2(\omega)},
\end{align}
where $\xi$ is the normal-state single-particle energy measured with respect to the Fermi level, $\Lambda\sim\omega_D$ is a cutoff for pairing interactions, $N(\xi)$ is the normal-state density of states, and we denote by $\tilde{\Omega}(\omega)$ and $\tilde{\Delta}(\omega)$ the zero-temperature versions of $\tilde{\omega}_n$ and $\tilde{\Delta}_n$, respectively. In the absence of disorder $\tilde{\Omega}(\omega)$ and $\tilde{\Delta}(\omega)$ reduce to $\omega$ and $\Delta$, respectively. One can study the ratio between the disordered and clean pair amplitudes,
\begin{align}\label{ratio}
\frac{\Delta_\text{eff}}{\Delta}&=\frac{\int\frac{d\omega}{2\pi}\int_{-\Lambda}^\Lambda d\xi\,N(\xi)\frac{\tilde{\Delta}(\omega)}{\xi^2+\tilde{\Omega}^2(\omega)+\tilde{\Delta}^2(\omega)}}{\int\frac{d\omega}{2\pi}\int_{-\Lambda}^\Lambda d\xi\,N(\xi)\frac{\Delta}{\xi^2+\omega^2+\Delta^2}}\nn\\
&=\frac{\int\frac{d\omega}{2\pi}\int_{-\Lambda}^\Lambda d\xi\,N(\xi)\frac{\tilde{\Delta}(\omega)}{\xi^2+\tilde{\Omega}^2(\omega)+\tilde{\Delta}^2(\omega)}}{\frac{1}{2}\int_{-\Lambda}^\Lambda d\xi\,N(\xi)\frac{\Delta}{\sqrt{\xi^2+\Delta^2}}},
\end{align}
which goes to one in the limit of vanishing disorder.

Let us first focus on the highly doped limit $\mu\gg\Lambda$. In this case the denominator of Eq.~(\ref{ratio}) is
\begin{align}
&\frac{1}{2}\int_{-\Lambda}^\Lambda d\xi\,N(\xi)\frac{\Delta}{\sqrt{\xi^2+\Delta^2}}\approx\frac{N(0)}{2}\int_{-\Lambda}^\Lambda d\xi\frac{\Delta}{\sqrt{\xi^2+\Delta^2}}\nn\\
&\hspace{20mm}=N(0)\Delta\ln\left[\frac{\Lambda}{\Delta}+\sqrt{\left(\frac{\Lambda}{\Delta}\right)^2+1}\right],
\end{align}
while the numerator is
\begin{align}
&\int\frac{d\omega}{2\pi}\int_{-\Lambda}^\Lambda d\xi\,N(\xi)\frac{\tilde{\Delta}(\omega)}{\xi^2+\tilde{\Omega}^2(\omega)+\tilde{\Delta}^2(\omega)}\nn\\
&\hspace{10mm}\approx N(0)\int\frac{d\omega}{2\pi}\int_{-\Lambda}^\Lambda d\xi\frac{\tilde{\Delta}(\omega)}{\xi^2+\tilde{\Omega}^2(\omega)+\tilde{\Delta}^2(\omega)}\nonumber\\
&\hspace{10mm}=\frac{N(0)}{\pi}\int d\omega\frac{\tilde{\Delta}(\omega)}{\sqrt{\tilde{\Omega}^2(\omega)+\tilde{\Delta}^2(\omega)}}\nn\\
&\hspace{25mm}\times\tan^{-1}\left(\frac{\Lambda}{\sqrt{\tilde{\Omega}^2(\omega)+\tilde{\Delta}^2(\omega)}}\right).
\end{align}
Taking the ratio, we obtain
\begin{align}\label{ratioLargeMu}
\frac{\Delta_\text{eff}(\Gamma)}{\Delta}&=\frac{1}{\pi\Delta\ln\left[\frac{\Lambda}{\Delta}+\sqrt{\left(\frac{\Lambda}{\Delta}\right)^2+1}\right]}
\nn\\
&\times\int d\omega\frac{\tilde{\Delta}(\omega)}{\sqrt{\tilde{\Omega}^2(\omega)+\tilde{\Delta}^2(\omega)}}\nn\\
&\times\tan^{-1}\left(\frac{\Lambda}{\sqrt{\tilde{\Omega}^2(\omega)+\tilde{\Delta}^2(\omega)}}\right),
\end{align}
where $\Gamma$ is the disorder strength. Solving the zero-temperature self-consistency conditions
\begin{align}
\tilde{\Omega}(\omega)&=\omega+\frac{\Gamma}{2}\frac{\tilde{\Omega}(\omega)}{\sqrt{\tilde{\Omega}^2(\omega)+\tilde{\Delta}^2(\omega)}},\label{ZeroT_SC1}\\
\tilde{\Delta}(\omega)&=\Delta-\frac{\Gamma}{2}\frac{\tilde{\Delta}(\omega)}{\sqrt{\tilde{\Omega}^2(\omega)+\tilde{\Delta}^2(\omega)}},\label{ZeroT_SC2}
\end{align}
and performing the integral over all frequencies in Eq.~(\ref{ratioLargeMu}), we obtain Fig.~\ref{fig:Delta_eff}(a). In practice, for every value of $\omega$ we solve the self-consistency conditions numerically by iteration, i.e.,
\begin{align}
\tilde{\Omega}^{(k+1)}(\omega)&=\omega+\frac{\Gamma}{2}\frac{\tilde{\Omega}^{(k)}(\omega)}{\sqrt{\tilde{\Omega}^{(k)}(\omega)^2+\tilde{\Delta}^{(k)}(\omega)^2}},\label{iter1}\\
\tilde{\Delta}^{(k+1)}(\omega)&=\Delta-\frac{\Gamma}{2}\frac{\tilde{\Delta}^{(k)}(\omega)}{\sqrt{\tilde{\Omega}^{(k)}(\omega)^2+\tilde{\Delta}^{(k)}(\omega)^2}},\label{iter2}
\end{align}
for $k=0,1,2,\ldots$, with the initial conditions $\tilde{\Omega}^{(0)}(\omega)=\omega$ and $\tilde{\Delta}^{(0)}(\omega)=\Delta$. The procedure is repeated until two successive iterations produce values that differ by less than a fixed tolerance. The frequency integral in Eq.~(\ref{ratioLargeMu}) is then also performed numerically.

We now turn to the undoped limit $\mu=0$. In this case the denominator of Eq.~(\ref{ratio}) is
\begin{align}
\frac{1}{2}\int_{-\Lambda}^\Lambda d\xi\,N(\xi)\frac{\Delta}{\sqrt{\xi^2+\Delta^2}}&=\frac{1}{2}\int_{-\Lambda}^\Lambda d\epsilon\,\frac{|\epsilon|}{2\pi v_F^2}\frac{\Delta}{\sqrt{\epsilon^2+\Delta^2}}\nn\\
&=\frac{\Delta^2}{2\pi v_F^2}\left(\sqrt{\left(\frac{\Lambda}{\Delta}\right)^2+1}-1\right),
\end{align}
and the numerator is
\begin{align}
&\int\frac{d\omega}{2\pi}\int_{-\Lambda}^\Lambda d\xi\,N(\xi)\frac{\tilde{\Delta}(\omega)}{\xi^2+\tilde{\Omega}^2(\omega)+\tilde{\Delta}^2(\omega)}\nn\\
&\hspace{10mm}=\int\frac{d\omega}{2\pi}\int_{-\Lambda}^\Lambda d\epsilon\,\frac{|\epsilon|}{2\pi v_F^2}\frac{\tilde{\Delta}(\omega)}{\epsilon^2+\tilde{\Omega}^2(\omega)+\tilde{\Delta}^2(\omega)}\nonumber\\
&\hspace{10mm}=\frac{1}{2\pi v_F^2}\int \frac{d\omega}{2\pi}\tilde{\Delta}(\omega)\ln\left(1+\frac{\Lambda^2}{\tilde{\Omega}^2(\omega)+\tilde{\Delta}^2(\omega)}\right),
\end{align}
thus the dimensionless ratio of amplitudes is
\begin{align}\label{ratioDeffMu0}
\frac{\Delta_\text{eff}(\gamma)}{\Delta}&=\frac{1}{\Delta^2\left(\sqrt{\left(\frac{\Lambda}{\Delta}\right)^2+1}-1\right)}\nn\\
&\times\int\frac{d\omega}{2\pi}\tilde{\Delta}(\omega)\ln\left(1+\frac{\Lambda^2}{\tilde{\Omega}^2(\omega)+\tilde{\Delta}^2(\omega)}\right).
\end{align}
In the undoped limit the self-consistency conditions for $\tilde{\Omega}(\omega)$ and $\tilde{\Delta}(\omega)$ are
\begin{align}
\tilde{\Omega}(\omega)&=\omega+\frac{\gamma\tilde{\Omega}(\omega)}{2}\ln\left(1+\frac{\Lambda^2}{\tilde{\Omega}^2(\omega)+\tilde{\Delta}^2(\omega)}\right),\\
\tilde{\Delta}(\omega)&=\Delta-\frac{\gamma\tilde{\Delta}(\omega)}{2}\ln\left(1+\frac{\Lambda^2}{\tilde{\Omega}^2(\omega)+\tilde{\Delta}^2(\omega)}\right),
\end{align}
which we solve by iteration as in Eq.~(\ref{iter1})-(\ref{iter2}). Performing the integral in Eq.~(\ref{ratioDeffMu0}) numerically, we obtain the result plotted in Fig.~\ref{fig:Delta_eff}(b).

\bibliography{gaplessSC}

\end{document}